\documentclass[prd,preprint,floatfix,nofootinbib,preprintnumbers,superscriptaddress,showkeys]{revtex4} 
\pdfoutput=1
\usepackage[caption=false]{subfig}
\usepackage{amsmath}
\usepackage{amsfonts}
\usepackage{amssymb}
\usepackage{float}
\usepackage{color}
\usepackage{graphicx}
\usepackage{graphics}
\usepackage{hyperref}
\usepackage{adjustbox,array}
\usepackage{enumitem} 
\hypersetup{}
\usepackage[utf8x]{inputenc} 
\usepackage{multirow}
\usepackage{booktabs}
\usepackage{mathrsfs}
\usepackage{diagbox}
\usepackage[english]{babel}
\usepackage[autostyle]{csquotes}

\graphicspath{{./figures/}}

\definecolor{rosso}{cmyk}{0,1,1,0.4}
\definecolor{rossos}{cmyk}{0,1,1,0.55}
\definecolor{rossoc}{cmyk}{0,1,1,0.2}
\definecolor{blu}{cmyk}{1,1,0,0.3}
\definecolor{blus}{cmyk}{1,1,0,0.6}
\definecolor{bluc}{cmyk}{1,1,0,0.1}
\definecolor{verde}{cmyk}{0.92,0,0.59,0.25}
\definecolor{verdec}{cmyk}{0.92,0,0.59,0.15}
\definecolor{verdes}{cmyk}{0.92,0,0.59,0.4}

\AtBeginDocument{
\heavyrulewidth=.08em
\lightrulewidth=.05em
\cmidrulewidth=.03em
\belowrulesep=.65ex
\belowbottomsep=0pt
\aboverulesep=.4ex
\abovetopsep=0pt
\cmidrulesep=\doublerulesep
\cmidrulekern=.5em
\defaultaddspace=.5em
}

\hypersetup{
 colorlinks=true,
 linkcolor=verdec,
 urlcolor=verdec,
 citecolor=verdec
}

\newcommand{\lsim}{\lesssim}

\newcommand{\lf}{\left(}
\newcommand{\ri}{\right)}

\newcommand{\nn}{\nonumber}

\renewcommand{\lg}{\mathscr{L}} 

\newcommand{\mco}{\mathcal{O}}

\newcommand{\mcs}{\mathcal{S}}

\newcommand{\br}{\text{Br}}

\newcommand{\lumtot}{\mathcal{L}_{\rm tot}}

\newcommand{\pb}{{\;{\rm pb}}}

\newcommand{\iab}{{\;{\rm ab}^{-1}}}

\newcommand{\mev}{{\;{\rm MeV}}}
\newcommand{\gev}{{\;{\rm GeV}}}
\newcommand{\tev}{{\;{\rm TeV}}}

\newcommand{\beq}{\begin{equation}}
\newcommand{\eeq}{\end{equation}}
\newcommand{\bea}{\begin{eqnarray}}
\newcommand{\eea}{\end{eqnarray}}
\newcommand{\barr}{\begin{array}}
\newcommand{\earr}{\end{array}}
\newcommand{\bc}{\begin{center}}
\newcommand{\ec}{\end{center}}
\newcommand{\bit}{\begin{itemize}}
\newcommand{\eit}{\end{itemize}}
\newcommand{\ben}{\begin{enumerate}}
\newcommand{\een}{\end{enumerate}}
\newcommand{\f}{\frac}

\newcommand{\al}{\alpha}

\newcommand{\Dt}{\Delta}

\newcommand{\sg}{\sigma}

\newcommand{\ves}{\varepsilon}

\newcommand{\gm}{\gamma}

\newcommand{\zd}{Z_{\rm D}}
\newcommand{\zdz}{Z_{\rm D,0}}
\newcommand{\mzd}{M_{Z_{\rm D}}}
\newcommand{\dmre}{\Delta m_{\rm recoil}}
\newcommand{\dmee}{\Delta m_{ee}}
\newcommand{\mre}{m_{\rm recoil}}
\newcommand{\mee}{m_{ee}}

\newcommand{\dtm}{\delta_m}

\newcommand{\ee}      {{e^+ e^-}}
\newcommand{\mmu}      {{\mu^+ \mu^-}}
\newcommand{\nnu}      {\nu\bar{\nu}}
\newcommand{\ttau}      {{\tau^+\tau^-}} 

\newcommand{\ttop}      {{t\bar{t}}}

\newcommand{\ww}      {{W^+ W^-}}

\newcommand{\met}      {{E_T^{\rm miss}}}

\newcommand{\pt}      {p_T}

\newcommand{\cw}      {c_{\rm W}}
\newcommand{\sw}      {s_{\rm W}}
\newcommand{\tw}      {t_{\rm W}}

\newcommand{\ca}      {c_\alpha}
\newcommand{\sa}      {s_\alpha}

\definecolor{mint}{rgb}{0.24, 0.71, 0.54}

\begin{document}
\preprint{KIAS-G099201}

\title{\color{verdes} Probing a Heavy Dark $Z$ Boson at Multi-TeV Muon Colliders: Leveraging the Optimized Recoil Mass Technique
} 
\author{Kingman Cheung}
\email{cheung@phys.nthu.edu.tw}
\address{Department of Physics, Konkuk University, Seoul 05029, Republic of Korea} 
\address{Department of Physics, National Tsing Hwa University, Hsinchu 300, Taiwan}
\address{Center for Theory and Computation, National Tsing Hua University,
  Hsinchu 300, Taiwan}
\author{Jinheung Kim}
\email{jhkim1216@kias.re.kr}
\address{School of Physics, Korea Institute for Advanced Study, Seoul 02455, Republic of Korea.} 
\author{Soojin Lee}
\email{soojinlee957@gmail.com}
\address{Department of Physics, Konkuk University, Seoul 05029, Republic of Korea}
\author{Prasenjit Sanyal}
\email{prasenjit.sanyal01@gmail.com}
\address{Department of Physics, Konkuk University, Seoul 05029, Republic of Korea}
\author{Jeonghyeon Song}
\email{jhsong@konkuk.ac.kr}
\address{Department of Physics, Konkuk University, Seoul 05029, Republic of Korea}

\begin{abstract} 
We investigate the discovery potential of multi-TeV muon colliders for a heavy dark $Z$ boson ($Z_{\rm D}$) with a mass above 1 TeV  through the associated production channel $\mu^+\mu^- \to Z_{\rm D}\gamma$. This process enables precise $M_{Z_{\rm D}}$ reconstruction using the photon recoil mass ($\mre$). Focusing on the $Z_{\rm D} \to jjX$ and $Z_{\rm D} \to e^+e^-$ decay modes, we present strategies for achieving high sensitivity to the kinetic mixing parameter $\varepsilon$ at 3, 6, and 10 TeV muon colliders with integrated luminosities of 1, 4, and 10 ab$^{-1}$ respectively, assuming $Z_{\rm D}$ decays exclusively into Standard Model particles. A key innovation is our optimized implementation of $M_{Z_{\rm D}}$-dependent cuts on $m_{\rm recoil}$, which accounts for the energy-dependent detector response. For heavier $Z_{\rm D}$, the associated photon becomes less energetic, leading to better photon energy resolution and thus enabling more stringent $m_{\rm recoil}$ cuts. This approach enhances $\varepsilon$ sensitivity for heavier $Z_{\rm D}$. Conversely, for lighter $Z_{\rm D}$, the lower-energy electron pair from $Z_{\rm D} \to e^+e^-$ enables tighter cuts on the invariant mass of the electron pair ($m_{ee}$), providing better sensitivity in the lighter mass regime. Combining these complementary $m_{\rm recoil}$- and $m_{ee}$-based selections with both $jjX$ and $e^+e^-$ channels, we achieve $\varepsilon$ sensitivity down to $\mathcal{O}\left(10^{-3}\right)$ as $M_{Z_{\rm D}}$ approaches $\sqrt{s}$, substantially surpassing the reach of a 100 TeV proton-proton collider. Even if $Z_{\rm D}$ decays into dark-sector particles, the recoil mass method remains effective, establishing muon colliders as powerful facilities for exploring heavy dark sectors.
\end{abstract}

\vspace{1cm}
\keywords{Dark Sector, Beyond the Standard Model, Muon Collider}

\maketitle
\tableofcontents
\flushbottom 
\section{Introduction}

Our universe is described by the Standard Model (SM) of particle physics, which has been thoroughly tested and confirmed, culminating in the landmark discovery of the Higgs boson at the Large Hadron Collider (LHC) in 2012~\cite{ATLAS:2012yve,CMS:2012qbp}. Despite this achievement, major unresolved puzzles persist: the fundamental properties of dark matter and dark energy, the origin of neutrino masses, and the matter-antimatter asymmetry. These profound questions strongly suggest that the SM is incomplete, motivating the search for new particles and interactions predicted by beyond the Standard Model (BSM) theories. High-energy collider experiments offer a controlled and precisely instrumented environment for these investigations. Yet, despite extensive efforts, the LHC has not uncovered direct evidence of BSM particles, underscoring the complexity of such probes.

In light of this challenge, a new paradigm emerged: the dark sector. Instead of assuming that new particles carry SM gauge charges, this scenario envisions them residing in a separate sector that interacts with our visible world only through a portal. One of the most compelling and minimal portals is realized by a dark $U(1)$ gauge boson $\zd$, which couples to the SM through kinetic mixing with a SM Abelian gauge boson~\cite{Okun:1982xi,Holdom:1985ag,Fayet:1990wx,Fabbrichesi:2020wbt}. This approach minimally extends the gauge principle and introduces a dimension-four mixing operator. If the dark sector particles are heavier than $\zd$, ensuring $\zd$ decays exclusively into SM particles, then the $\zd$ phenomenology depends only on its mass $\mzd$ and the kinetic mixing parameter $\varepsilon$.

Extensive experimental efforts have tightly constrained the lighter $\zd$ mass range using various fixed-target, beam-dump, and collider experiments, as well as astrophysical observations. For $1\mev \lsim \mzd \lsim 30\mev$, a wide range of $\varepsilon > \mathcal{O}(10^{-9})$ is excluded   at the $2\sigma$ level by the supernovae bound~\cite{Chang:2016ntp}, beam dump experiments E774~\cite{Bross:1989mp}, CHARM~\cite{Gninenko:2012eq}, $\nu$-Cal~\cite{Blumlein:2011mv,Blumlein:2013cua}, and E141~\cite{Riordan:1987aw}, as well as NA64(e)~\cite{NA64:eplus}, NA48/2~\cite{Batley:2015lha}, and $(g-2)_e$~\cite{Pospelov:2008zw}. In the mass window $30\mev \lsim \mzd \lsim 550\mev$, numerous existing and proposed experiments including SHiP~\cite{Alekhin:2015byh,Anelli:2015pba}, FASER/FASER2~\cite{Feng:2017uoz,Cheung:2022kjd}, NA64(e)$^{++}$~\cite{NA64:eplus}, SeaQuest~\cite{Berlin:2018pwi}, HPS~\cite{Adrian:2018scb}, MESA~\cite{Doria:2019sux,Doria:2018sfx}, and LHeC/FCC-eh~\cite{DOnofrio:2019dcp} are expected to probe $\varepsilon$ to $\mathcal{O}(10^{-8})$. For $550\mev \lsim \mzd \lsim 70\gev$, current bounds from LHCb~\cite{Aaij:2019bvg}, BaBar~\cite{Lees:2014xha}, and CMS~\cite{CMS:2019kiy}, as well as projected sensitivities from Belle-II~\cite{Kou:2018nap} and FCC-ee~\cite{Karliner:2015tga}, indicate that $\varepsilon$ can be probed up to $\mathcal{O}(10^{-3})$. Finally, in the regime $70\gev \lsim \mzd \lsim 1\tev$, the model is constrained by the Drell-Yan process $pp \to \zd \to \ell^+\ell^-$ at the LHC and by indirect effects on electroweak precision observables~\cite{Curtin:2014cca}, reaching $\varepsilon \sim \mathcal{O}(10^{-2})$.

However, for a heavy $\zd$ with a mass above $1\tev$, constraints become far weaker. Existing studies have predominantly focused on proton-proton ($pp$) colliders via the Drell-Yan process. At the HL-LHC, with an integrated luminosity of $3\iab$, the projected sensitivity reaches $\varepsilon \sim 10^{-2}$ and $0.1$ for $\mzd = 1\tev$ and $2.4\tev$, respectively. A future 100 TeV $pp$ collider with $3\iab$ could potentially improve these limits to $\varepsilon \sim 4\times 10^{-3}$ and $3\times 10^{-2}$ for the same mass points. Notably, as $\mzd$ increases, the achievable $\varepsilon$ limit becomes larger, making it increasingly challenging to explore very heavy $\zd$. Moreover, since a 100 TeV $pp$ collider remains a distant prospect, exploring alternative approaches to probe such heavy $\zd$ masses deserves serious consideration.

A multi-TeV muon collider (MuC)~\cite{Palmer:1996gs,Ankenbrandt:1999cta,Schulte:2021eyr} offers a promising solution. Capable of achieving multi-TeV center-of-mass (c.m.) energies in a relatively clean environment, the MuC can fully exploit its collision energy for BSM searches~\cite{Capdevilla:2020qel,Bandyopadhyay:2021pld, Sen:2021fha, Asadi:2021gah, Huang:2021nkl}. Previous studies have shown that MuCs excel at probing heavy new particles, surpassing hadron colliders in their kinematic reach. Recent progress in muon beam cooling~\cite{Antonelli:2013mmk,Antonelli:2015nla} and in managing beam-induced backgrounds (BIBs)~\cite{Collamati:2021sbv,Ally:2022rgk} has made the MuC increasingly realistic. Although indirect effects of a dark $Z$ at a MuC have been investigated for moderate masses below about 1.5 TeV~\cite{Hosseini:2022urq}, direct search strategies for a heavy $\zd$ at multi-TeV MuCs remain largely unexplored.

One of the most powerful techniques to enhance the discovery potential of a MuC for the heavy $\zd$ is the recoil mass ($\mre$) method, which takes full advantage of the known initial-state kinematics in lepton collisions~\cite{Chakrabarty:2014pja,Draper:2018ljh,Han:2020pif,Zhu:2022lzv,Sha:2022bkt,Dasgupta:2023zrh,Chen:2022yiu,Denizli:2023rqe,Forslund:2023reu,Jiang:2024wwa,Li:2024joa,Barik:2024kwv}. In particular, the associated production channel $\mu^+\mu^- \to \zd\gamma$ is promising: the known c.m.~energy and the measured photon energy $E_{\gamma}$ determine $\mzd$ via $\mre^2 = s - 2\sqrt{s}E_{\gamma}$, without assumptions about how $\zd$ decays. This recoil mass technique cannot be employed at hadron colliders due to unknown initial parton energies. Furthermore, the $\zd \to e^+e^-$ decay mode provides a complementary mass reconstruction method via the $e^+e^-$ invariant mass $\mee$.

While the recoil mass method is well-established, previous analyses typically employed universal, mass-independent cuts on both $\mre$ and $\mee$, without accounting for energy resolution effects at the detector level. In this work, we introduce a more refined approach featuring $\mzd$- and $\sqrt{s}$-dependent mass cuts. Since a heavier $\zd$ is accompanied by a less energetic photon, leading to better photon energy resolution~\cite{Chakrabarty:2014pja}, we can implement more stringent $\mre$ cuts for heavier $\zd$. This innovation significantly enhances the $\varepsilon$ sensitivity in the heavier mass regime. Meanwhile, for the $\zd \to e^+e^-$ channel, a lighter $\zd$ produces a lower-energy electron pair, enabling tighter $\mee$ cuts for lighter $\zd$. Therefore, the interplay of $\mre$- and $\mee$-based selections  ensures robust coverage across a broad $\mzd$ range. Combined with the increasing cross section of $\mu^+\mu^- \to \zd\gamma$ as $\mzd$ approaches $\sqrt{s}$, our method opens a powerful new channel for probing heavy $\zd$, marking our primary contribution.

The organization of this paper is as follows. In Section~\ref{sec-review}, we briefly review the theory of kinetically mixed $U(1)_D$. Drawing on the characteristic features of the heavy dark $Z$, we identify the golden channels to probe $\zd$ at a multi-TeV MuC. Section~\ref{sec-recoil-mass} discusses the  mass reconstruction technique  in detail, including the introduction of $\mzd$-dependent resolutions $\dmre$ and $\dmee$. In Section~\ref{sec-analysis}, we present a comprehensive signal-to-background analysis, applying mass-dependent cuts for both $\mre$ and $\mee$ and demonstrating the resulting improvements in $\varepsilon$ sensitivity. We then offer sensitivity projections for various MuC energies, highlighting the complementarity of the $jjX$ and $e^+e^-$ decay modes across different $\mzd$ regimes, and compare our results with other future $pp$  colliders. Finally, we conclude in Section~\ref{sec-conclusion}, summarizing our findings and discussing the implications for future dark sector searches.

\section{Theory of Heavy Dark $Z$ and Identification of Golden Discovery Channels at Muon Colliders}
\label{sec-review}

In this section, we briefly review the theoretical framework of the dark $Z$ boson ($\zd$), which emerges from kinetic mixing with the SM hypercharge gauge field $\tilde B_{\mu}$. The relevant Lagrangian terms are given by~\cite{Fabbrichesi:2020wbt}: 
\beq\label{eq-Lag}
\lg \supset -\frac{1}{4} \,\tilde B_{\mu\nu}\, \tilde B^{\mu\nu} 
- \frac{1}{4} \,\tilde Z_{\rm D\mu\nu}\, \tilde{Z}_\text{D}^{\mu\nu}  
- \f{1}{2}\,\f{\varepsilon}{\cw} \,\tilde Z_ {\rm D\mu\nu}\,\tilde B^{\mu\nu} 
+ \f{1}{2}\, M_{D,0}^2\, \tilde{Z}_\text{D}^\mu \, \tilde Z_{\rm D\mu}\, ,
\eeq
where $\tilde B^{\mu\nu}
=\partial^\mu \tilde B^{\nu} - \partial^\nu \tilde B^{\mu}$ and 
$\tilde{Z}_{\rm D}^{\mu\nu} =\partial^\mu \tilde{Z}_{\rm D}^\nu - \partial^\nu \tilde{Z}_{\rm D}^\mu$,
with $\cw = \cos\theta_\text{W}$ being the cosine of the Weinberg angle $\theta_\text{W}$.
For convenience, we define $s_x = \sin x$, $c_x = \cos x$, and $t_x = \tan x$.
In order to probe parameter regions that remain relatively unexplored by current and future collider experiments, we focus on a heavy dark $Z$ scenario with small kinetic mixing: 
\bea
\label{eq-mass-range}
\mzd \in [1,10]\tev,\quad \varepsilon < 10^{-1}.
\eea 

The mass term in \autoref{eq-Lag} can originate from a Stueckelberg mechanism~\cite{Ruegg:2003ps} or a dark Higgs mechanism~\cite{Curtin:2014cca,Fabbrichesi:2020wbt}, but its underlying source does not critically impact the phenomenology at high-energy colliders. For heavy masses satisfying \autoref{eq-mass-range}, we introduce a small mass ratio parameter: 
\bea
\label{eq-dtm}
\dtm = \frac{m_{Z,0}}{M_{D,0}} \ll 1,
\eea
where $m_{Z,0} = g v /(2 \cw) $ with $v \simeq 246\gev$. 

The field redefinition of
\bea
\label{eq-field-re}
\left(\begin {array}{c} \zdz^\mu \\ B^\mu \end{array}\right) = 
       \left(\begin {array}{cc} \sqrt{1-\frac{\ves ^2}{\cw^2}} & 0 \\ \frac{\ves}{\cw} & 1 \end{array}\right)
      \left(\begin {array}{c} \tilde{ Z}_{\rm D}^\mu \\ \tilde{B}^\mu \end{array}\right)
\eea
diagonalizes the kinetic terms of the gauge bosons.
After the electroweak symmetry breaking, the mass-squared matrix 
in the basis of $(A^\mu, Z_0^\mu, \zdz^\mu)$ becomes
\beq
\label{eq-mass2-matrix}
\mathcal{M}^2_V = M_{D,0}^2 
\left(\begin {array}{ccc} 
0  & 0 &  0 \\  
0 & \dtm^2 &  \tw \dtm^2 \ves \\
0 &  \tw \dtm^2 \ves& 1 \end{array}\right) + \mco\lf \dtm^2\ves^2 \ri ,
\eeq
clearly  showing a massless photon eigenstate.

The physical $Z$ and $\zd$ mass eigenstates are obtained through the mixing matrix:
\beq
\label{eq-ZZd-mixing}
\left(\begin {array}{c} Z \\ \zd \end{array}\right) = 
      \left(
      \begin {array}{rr} \ca\;\; &\sa \\ -\sa\;\; &\ca 
      \end{array}\right)
     \left(\begin {array}{c} Z_0 \\ \zdz\end{array}\right),
\eeq
where the mixing angle $\al$ is given by
\beq
\label{eq-alpha}
\al = -\tw \dtm^2 \ves + \mco\lf  \dtm^4 \ves \ri.
\eeq
The masses of the physical $Z$ and $\zd$ are
\begin{align}
\label{eq-mass}
m_Z & = m_{Z,0} \left[
1 - \f{1}{2} \,\tw^2 \dtm^2 \ves^2 + \mco \lf \dtm^4 \ves^2 \ri 
\right] ,
\\ \nn
\mzd &= M_{D,0} 
 \left[
1 + \f{1}{2} \,\tw^2 \dtm^2 \ves^2 + \mco \lf \dtm^4 \ves^2 \ri 
\right].
\end{align}
Thus, $Z$ and $\zd$ are canonically normalized and have well-defined masses. 

The interaction Lagrangian for $Z$ and $\zd$ couplings to SM fermions is
\bea
\label{eq-Lag-int}
-\lg_{\rm int}^{f}=  \f{g}{\cw}\,\hat{g}_{Z}^f Z^\mu \bar{f} \gm_\mu f 
+\ves \, \f{g}{\cw} \,\hat{g}_{\zd}^f \zd^\mu \bar{f} \gm_\mu f.
\eea
The normalized couplings are given by
\begin{align}
\label{eq-coupling}
\hat{g}_{Z}^f &= 
T_{3f} - Q_f \sw^2 - ( T_{3f}-Q_f  )\tw^2 \dtm^2 \ves^2 + \mco(\dtm^4 \ves^2),
 \\ \nn
\hat{g}_{\zd}^f &=  
  ( T_{3f}-Q_f  )\tw   + \mco(\dtm^2)
,
\end{align}
where $T_{3f}$ and $Q_f$ are the third component of weak isospin and the electric charge
for the fermion $f$, respectively.
Through the $Z$ and $\zd$ mixing in \autoref{eq-ZZd-mixing}, 
the dark $Z$ also couples to $W^+W^-$ and $Z H$, where $H$ is the 125 GeV Higgs boson. 

With these interaction vertices established, we now turn to the decay channels of $\zd$. To facilitate a fair comparison with existing LHC sensitivity limits, we do not consider $\zd$ decays into BSM states such as dark matter~\cite{Abdallah:2021npg,Barik:2024kwv}. Instead, we focus solely on decays into SM particles, following the same setup employed in the conventional LHC searches~\cite{ALEPH:2013dgf,ATLAS:2019erb,LHCb:2017trq,ATLAS:2017eiz,ATLAS:2018mrn,CMS:2018hnz,ATLAS:2017ptz,CMS:2021fyk,CMS:2018yxg,ATLAS:2018coo,ATLAS:2019nat,CMS:2019qem,D0:2010kuq,BaBar:2014zli}. This approach allows us to clearly demonstrate the improved $\ves$ sensitivity at a multi-TeV MuC.
 
For $\mzd>1\tev$, the dominant $\zd$ decays are into SM fermion pairs, including top quarks. However, the relative importance of specific decay channels differs significantly from that of the SM $Z$ boson. For instance, the ratios:
\begin{align}
\label{eq-nu-coupling-ratio}
\f{\hat{g}_{\zd}^{\nu_L}}{\hat{g}_{Z}^{\nu_L} }&= \tw \approx 0.548,
\\[5pt] \nn
\f{\hat{g}_{\zd}^{e_R}}{\hat{g}_{Z}^{e_R} }&= \f{1}{\sw\cw} \approx 2.37,
\quad  \f{\hat{g}_{\zd}^{e_L}}{\hat{g}_{Z}^{e_L} }= \f{\tw}{2\sw^2-1} \approx -1.02,
\end{align}
show that neutrino modes are relatively suppressed, whereas the decay into an electron-positron pair is enhanced.

The branching ratios remain nearly constant for $\mzd \geq 1\tev$: 
\begin{align}
\label{eq-br}
\sum_q \br (\zd \to q\bar{q}) &\approx 0.40, \quad ~\br (\zd \to \ttop)\approx 0.13,
\\ \nn
\sum_{\nu_i}\br (\zd \to \nu_i \bar{\nu}_i)&\approx 0.074, \quad \br (\zd \to e^+ e^-)\approx 0.12,
\\ \nn
\br (\zd \to \ww)&\approx 0.019, \quad
\br (\zd \to Z H)\approx 0.013,
\end{align}
where $q=u,d,c,s,b$, and $\nu_i=\nu_e,\nu_\mu, \nu_\tau$. For $\mzd>1\tev$, we have $\br(\zd \to e^+ e^-)=\br(\zd \to \mmu)=\br(\zd \to \ttau)$. We note that
the suppression of the neutrino decay mode, combined with the enhancement of the charged lepton decay  channel, distinguishes the heavy $\zd$ from the SM $Z$ boson.

In light of these decay patterns, we focus on two key modes at the MuC: 
\bea
\label{eq-decay-modes}
\zd \to jj+X,\quad \zd \to \ee.
\eea
The inclusive dijet mode combines multiple hadronic final states ($q\bar{q}$, $\ttau$, $\ttop$, and $\ww$), maximizing the signal event yield and enhancing sensitivity.
The $e^+e^-$ mode, on the other hand, provides a clean leptonic final state with high-precision momentum measurements and excellent invariant mass reconstruction. 
We focus on the $e^+e^-$ final state rather than the $\mu^+\mu^-$ channel, as the latter faces increased backgrounds at a multi-TeV MuC, notably from additional vector boson fusion (VBF) processes that produce muons in the final state.

We now examine the production channel of a heavy dark $Z$ at a multi-TeV MuC. Direct single $\zd$ production via an $s$-channel resonance is inefficient since it requires $\sqrt{s} \approx \mzd$. Although initial state radiation (ISR) can broaden the accessible mass range, this approach remains more limited at MuCs than at $e^+e^-$ colliders due to reduced ISR from muons. Instead, associated production with SM particles through $2\to2$ or $2\to3$ processes is more promising, as it enables exploration of a wide mass range with $\mzd < \sqrt{s}$ while providing additional hard SM particles that help suppress BIBs.

Motivated by these considerations, we examine three primary production channels:
\begin{align}
\label{eq-production-3ch}
\mmu &\to \zd\gm,
\\ \nn
\mmu &\to \zd \nu\bar{\nu},
\\ \nn
\mmu &\to \zd \mmu,
\end{align}
where $\nu$ includes all three flavors.

\begin{figure}[t]
	\centering
	\includegraphics[width=0.47\linewidth]{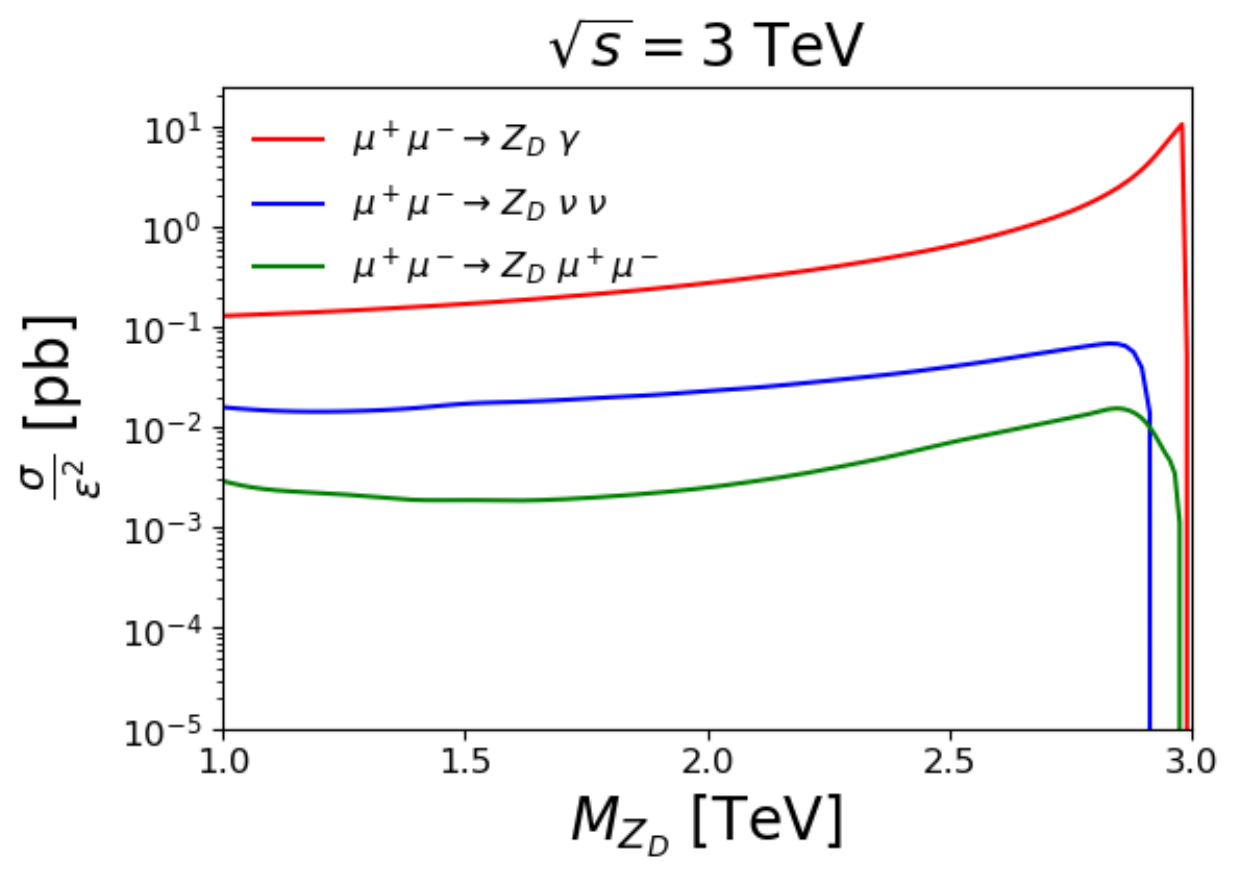}~~
	\includegraphics[width=0.47\linewidth]{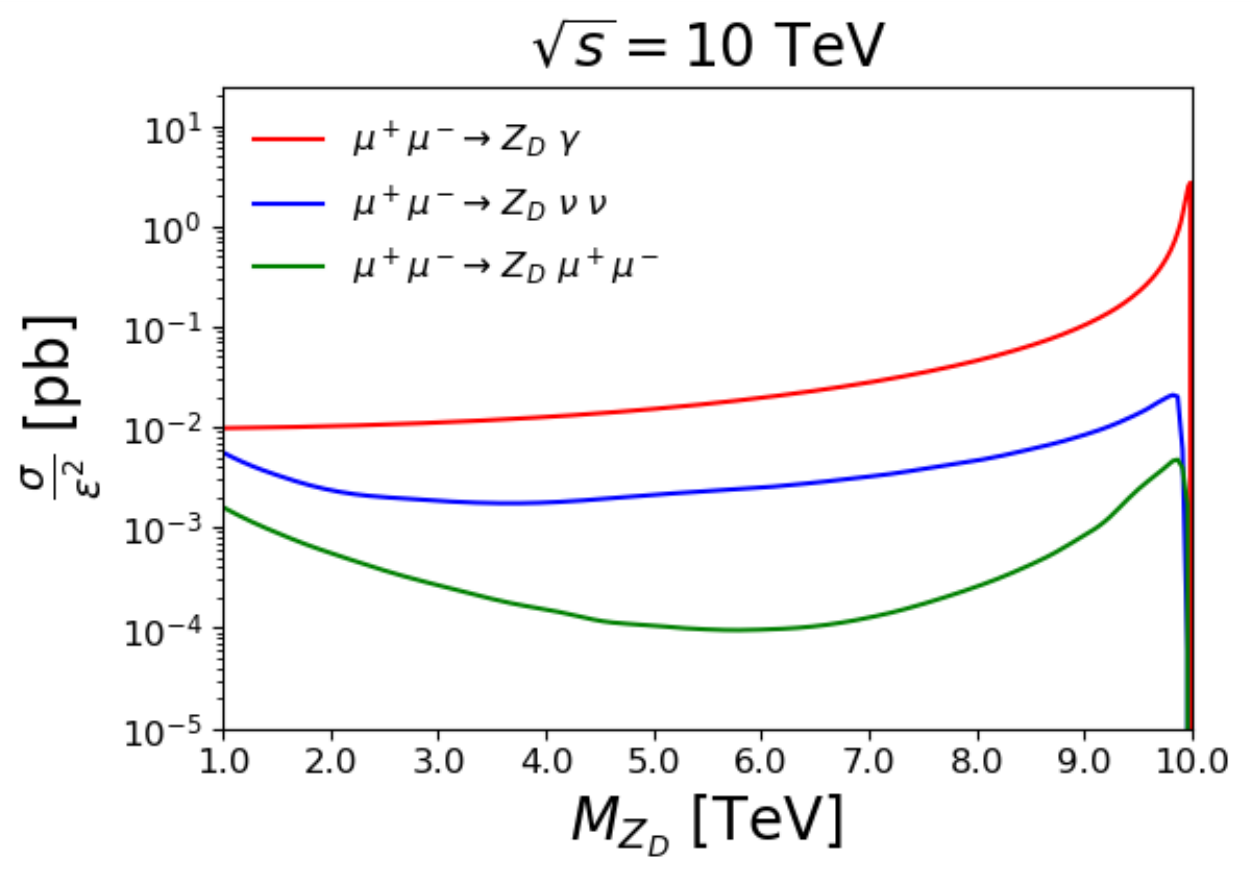}
	\caption{Cross sections of heavy dark $Z$ production as a function of $\mzd$ through $\mmu \to \zd\gm$ (red), $\mmu \to \zd \nu\bar{\nu}$ (blue), and $\mmu \to \zd \mmu$ (green) at MuCs with $\sqrt{s}=3 \tev$ (left) and $\sqrt{s}=10 \tev$ (right). The cross sections are normalized by $\ves^2$.}
	\label{fig-xsec}
\end{figure}

\autoref{fig-xsec} shows the parton-level cross sections (normalized by $\ves^2$) as a function of $\mzd$ at $\sqrt{s}=3 \tev$ (left) and $\sqrt{s}=10 \tev$ (right), obtained using {\small\sc MadGraph5-aMC@NLO}~\cite{Alwall:2011uj}. For the $\zd\gm$ and $\zd\mmu$ channels, we require $p_T^{\gm,\mu} \geq 10\gev$ and $|\eta_{\gm,\mu}|\leq 2.5$, while for the $\zd \nnu$ channel we impose $\met \geq 20\gev$. 

At both 3 TeV and 10 TeV MuC energies, the $\mmu \to \zd\gm$ channel typically yields the largest cross section, followed by $\mmu \to \zd \nnu$ and then $\mmu \to \zd \mmu$. As $\mzd$ approaches $\sqrt{s}$, the $\zd\gamma$ cross section rises sharply because the photon becomes softer, effectively reducing the process to a $2\to1$ configuration. However, as $\mzd$ nears $\sqrt{s}$, the available phase space is severely constrained by the $p_{T}^{\gamma} > 10\gev$ requirement, causing the cross section to drop precipitously.

This behavior contrasts with Drell-Yan process at $pp$ colliders, where higher $\mzd$ values result in significantly smaller cross sections due to the limited parton distribution functions at large momentum fractions. For instance, at a 100 TeV $pp$ collider, $\sigma(pp \to \zd)/\ves^2 =140,\,0.5,\,0.03\pb$ for $\mzd=1,\,5,\,10\tev$, respectively. 

Among the subdominant channels, $\mmu \to \zd \nnu$ benefits from three neutrino flavors and VBF contributions at high energies~\cite{Costantini:2020stv}. In contrast, $\mmu \to \zd \mmu$ is suppressed due to the absence of a $Z$-$Z$-$\zd$ vertex, which is required for charged-current VBF processes. Combining these insights on production and decay, we identify two ``golden'' discovery modes for heavy $\zd$ at the MuC: $\mmu \to \zd\gm$ followed by $\zd \to jj+X$ or $\zd \to \ee$. These channels provide the strongest prospect for discovering a heavy dark $Z$ at future multi-TeV MuCs.

\section{Precision Measurement of Heavy Dark $Z$ Mass at Muon Colliders}
\label{sec-recoil-mass}

Having identified two golden channels in the previous section, we now address a key component of our analysis: the precise measurement of the heavy $\zd$ mass. Accurate determination of $\mzd$ is essential for efficient background suppression, thereby forming the backbone of our analysis strategy.

A common approach to measuring $\mzd$ is to use the resonance peak in the invariant mass of the $\zd$ decay products. For the $\zd \to e^+ e^-$ channel, the invariant mass of the electron-positron pair ($m_{ee}$) directly provides this resonance. However, this method does not work for the inclusive dijet mode including $\zd \to t\bar{t}$, $\zd \to \tau^+\tau^-$, and $\zd \to \ww$.

Fortunately, the production channel $\mmu \to \zd\gm$ offers a powerful alternative: measuring the recoil mass $\mre$ of the photon. In the c.m.~frame, $\mre$ is defined by
\bea
\label{eq-recoil-mass-def}
\mre^2 =  s -2 \sqrt{s} E_\gm = \mzd^2.
\eea
Thus, the heavy dark $Z$ mass is uniquely determined by $\sqrt{s}$ and the photon energy. This recoil mass technique is a well-established method for BSM searches at lepton colliders.

However, many existing studies on mass reconstruction employ a fixed mass window cut (e.g., $|m_{\text{recoil},ee} - \mzd | < 10\gev$)~\cite{Han:2020pif}, regardless of $\mzd$. This simplistic approach overlooks critical detector effects. At the detector level, the widths of the $\mre$ and $\mee$ distributions vary significantly with both $\mzd$ and $\sqrt{s}$. This variation arises because the energy resolutions for the photon, electron, and positron strongly depend on their energies. Consequently, applying a universal 10 GeV mass window can be problematic: if the resolution exceeds 10 GeV, the cut may exclude too many signal events; conversely, if the resolution is smaller than 10 GeV,  the cut may be unnecessarily wide, allowing excessive background. Therefore, a more nuanced, $\mzd$-dependent approach is necessary to optimize signal selection and background rejection across the entire $\mzd$ range at a given $\sqrt{s}$.

\begin{figure}[t]
    \centering
    \includegraphics[width=0.9\textwidth]{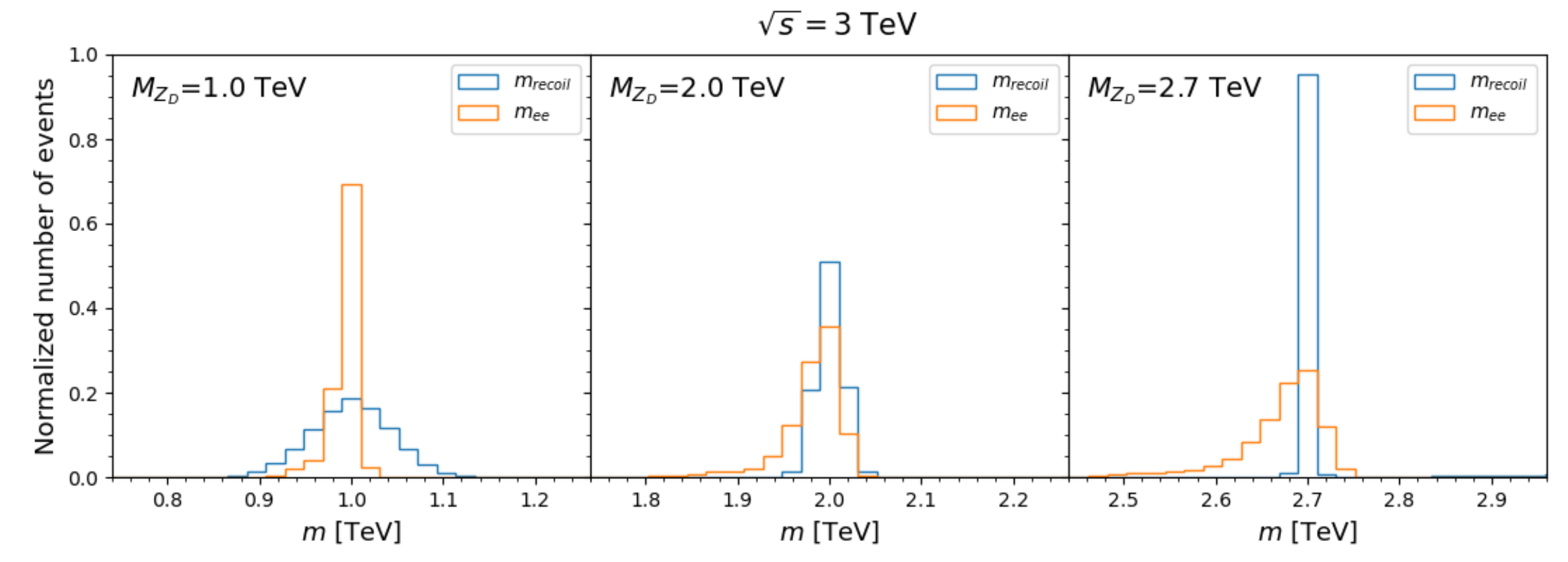}
    \\[5pt]
     \includegraphics[width=0.9\textwidth]{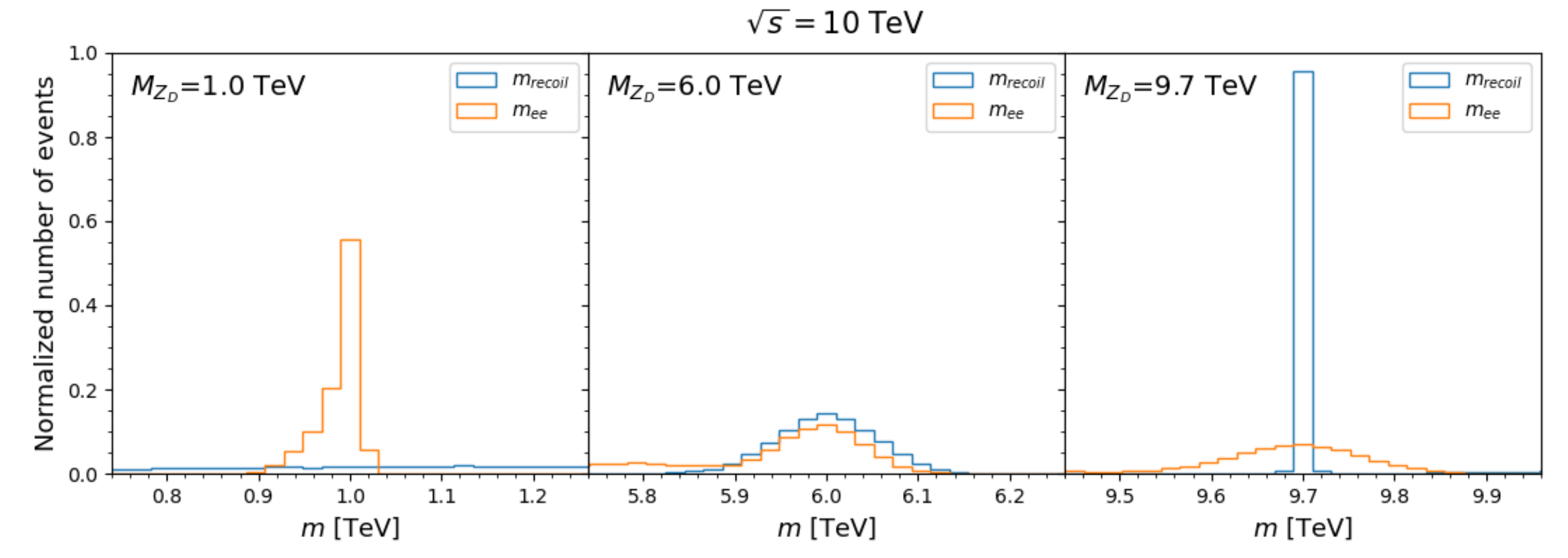}
  \caption{Normalized distributions of the recoil mass $\mre$ (blue) and
the $e^+e^-$ invariant mass $\mee$ (orange)
for $\mmu \to \zd(\to e^+e^-)\gm$ at $\sqrt{s}=3$ TeV (upper panels) and
$\sqrt{s}=10$ TeV (lower panels) MuC.
}
\label{fig-mre-mee}
\end{figure}

To illustrate this point, \autoref{fig-mre-mee} presents the normalized distributions of $\mre$ (blue) and $\mee$ (orange) at the detector level for $\mmu \to \zd(\to e^+e^-)\gm$ at $\sqrt{s}=3$ TeV (upper panels) and $\sqrt{s}=10$ TeV (lower panels), obtained using \textsc{Delphes} version 3.5.1~\cite{deFavereau:2013fsa} with the \texttt{delphes\_card\_MuonColliderDet.tcl} detector configuration. We observe that as $\mzd$ increases, the $\mre$ distribution becomes narrower, while the $\mee$ distribution becomes broader.

The behavior of the $\mre$ distribution is primarily attributed to the photon energy resolution $\Delta E_\gamma$ at the detector. Typically, $\Delta E_\gamma$ is expressed as:
\bea \label{eq-Eresolution-ECAL} 
\frac{\Delta E_\gamma}{E_\gamma} = \frac{a}{\sqrt{E_\gamma}} \oplus b \oplus \frac{c}{E_\gamma}, 
\eea
where $E_\gamma$ is measured in GeV, and $\oplus$ denotes addition in quadrature. In the \texttt{delphes\_card\_MuonColliderDet.tcl}, the parameters are set to $b=0.01$, $c=0$, and $a$ varies with pseudorapidity as follows: $a=0.156$ for $|\eta_\gamma| \leq 0.78$, $a=0.175$ for $0.78 \leq |\eta_\gamma| \leq 0.83$, and $a=0.151$ for $0.83 \leq |\eta_\gamma| \leq 2.5$. 

\autoref{fig-mre-mee} shows that, due to this energy resolution, the $\mre$ distribution is much broader for lighter $\mzd$ (which corresponds to higher-energy photons) and sharper for heavier $\mzd$ (which corresponds to lower-energy photons). For example, at a 10 TeV MuC, we find $\Dt E_\gm \approx 50\gev$ for $\mzd=1\tev$, which is compared to only $\Dt E_\gm \approx 2\gev$ for $\mzd=9.8\tev$. This significant improvement in resolution at higher masses strongly motivates using $\mzd$-dependent mass windows for $\mre$ to maximize signal sensitivity.

\begin{figure}[t]
    \centering
    \includegraphics[width=0.65\textwidth]{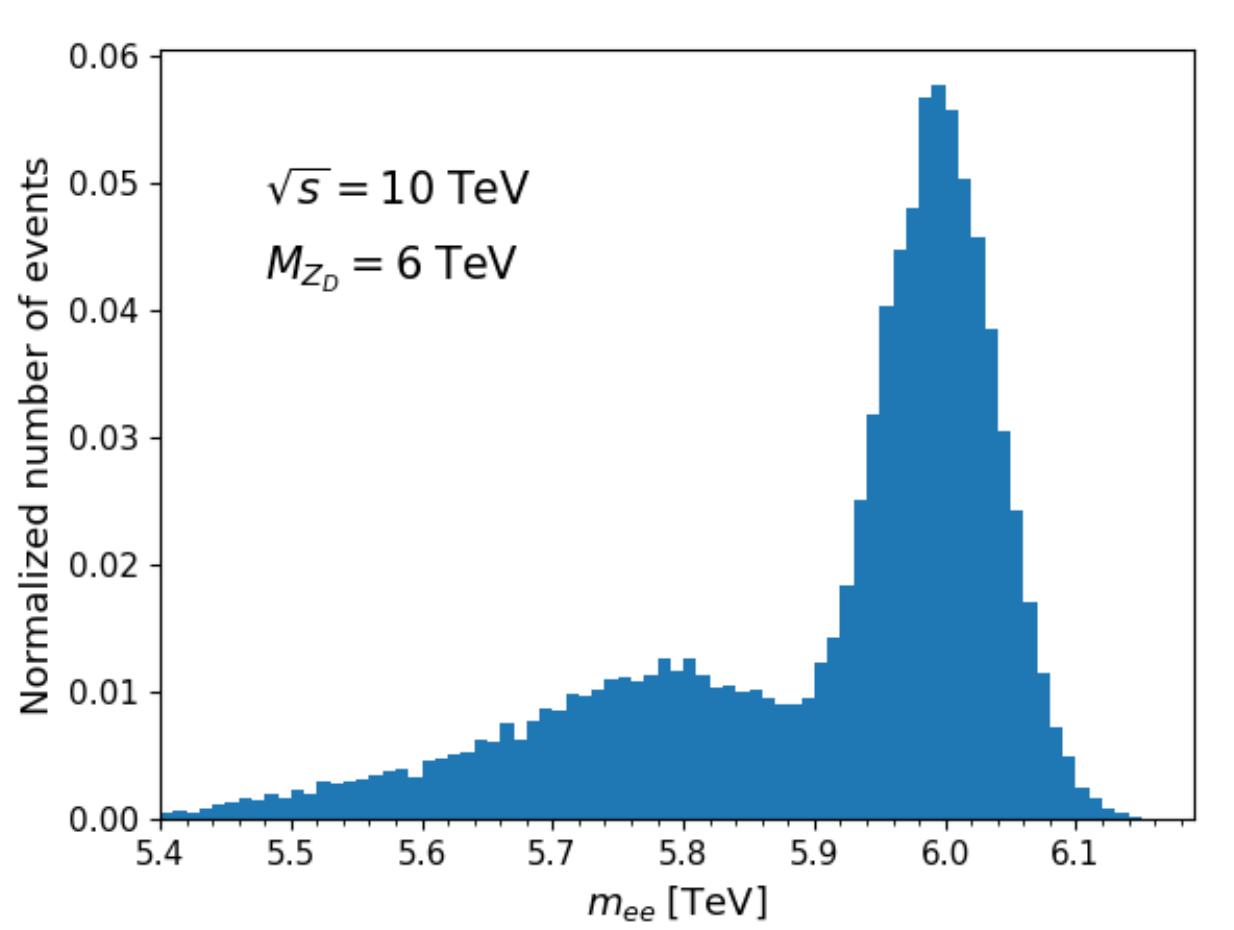}
  \caption{Normalized distributions of the $\mee$ invariant mass for $\mmu \to \zd\gm$ with $\mzd=6$ TeV at the 10 TeV MuC.
}
  \label{fig-mee-MZp6-10TeV}
\end{figure}

Interestingly, the $\mee$ distribution\footnote{The electron and photon energy resolution parameters are set to be identical in the \textsc{Delphes} card.} exhibits the opposite behavior, becoming broader for heavier $\mzd$ as illustrated in \autoref{fig-mre-mee}. Another distinctive feature is its increasingly asymmetric shape with a pronounced tail toward lower values as $\mzd$ increases. To demonstrate this asymmetry in detail, \autoref{fig-mee-MZp6-10TeV} shows the $\mee$ distribution for $\mzd=6\tev$ at a 10 TeV MuC. Several factors contribute to this asymmetry: the energy and momentum resolutions at the detector deteriorate for more energetic electrons and positrons; bremsstrahlung losses become more significant at higher particle energies, shifting the measured energies downward; and final state radiation (FSR) is particularly pronounced for electrons and positrons due to their low mass. Collectively, these effects cause the $\mee$ distribution to shift toward lower values and broaden with increasing $\mzd$.

These contrasting behaviors of the $\mre$ and $\mee$ distributions emphasize the importance of adopting different mass window cuts for each distribution, tailored to $\mzd$: 
\bea \label{eq-mass-window}
 \left| \mre - \mzd \right| < 2 \dmre, \quad \left| \mee - \mzd \right| < 2 \dmee.
\eea 
By optimizing $\dmre$ and $\dmee$ according to $\mzd$ and $\sqrt{s}$, we can improve the $\ves$ sensitivity across the full mass range.

To determine the optimal values of $\dmre$ and $\dmee$, we conducted detailed detector-level simulations at multiple benchmark points of $\mzd$ for 3 TeV, 6 TeV, and 10 TeV MuC. We generated parton-level events using \textsc{MadGraph5}, processed them through \textsc{Pythia} 8.307~\cite{Bierlich:2022pfr} for parton showering and hadronization, and finally employed \textsc{Delphes} for detector simulation. The resulting $\mre$ and $\mee$ distributions were then fitted with Gaussian functions to characterize their shapes.\footnote{For the asymmetric $\mee$ distribution, we restricted the Gaussian fit to a window of $\pm25\%$ around the true $\mzd$ in the signal event counts. This approach mitigates the impact of the asymmetric tail while accurately capturing the core of the distribution, a strategy justified by the alignment of the $\mee$ peak position with the true $\mzd$.}
For each $\mzd$ and $\sqrt{s}$, we set $\dmre$ and $\dmee$ to be the standard deviations obtained from the Gaussian fits of the $\mre$ and $\mee$ distributions, respectively.

\begin{figure}[t]
    \centering
    \includegraphics[width=0.65\textwidth]{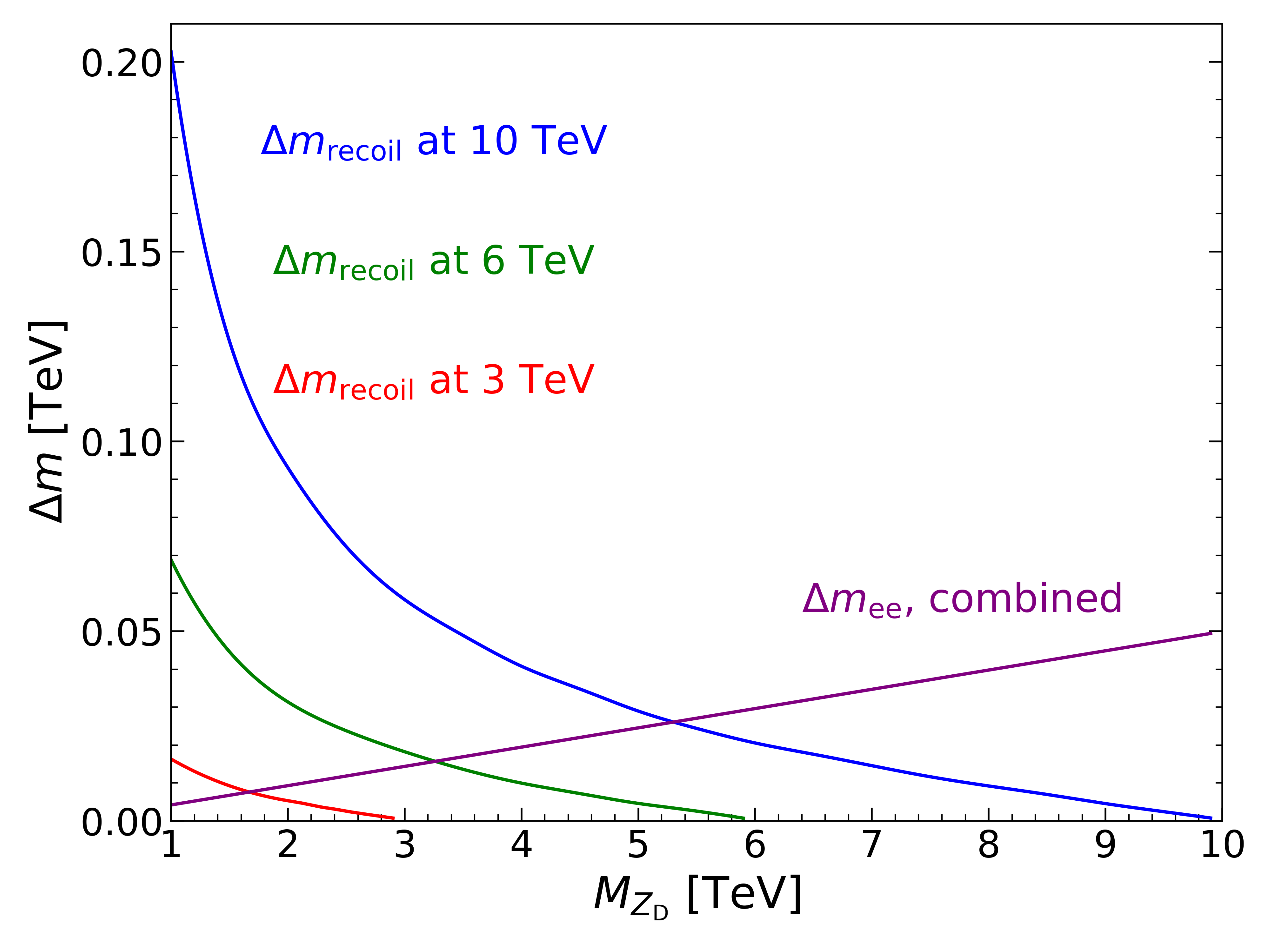}
  \caption{$\dmre$ and $\dmee$, the standard deviations obtained from Gaussian fits to the $\mre$ and $\mee$
  distributions,
  as a function of $\mzd$
  at the 3 TeV, 6 TeV, and 10 TeV MuC.
}
  \label{fig-Delta-Mass}
\end{figure}

\autoref{fig-Delta-Mass} displays the extracted $\dmre$ and $\dmee$ values as functions of $\mzd$ at $\sqrt{s}=3$ TeV, 6 TeV, and 10 TeV MuC. 
Since the $\dmee$ values are similar across the three $\sqrt{s}$ points within their kinematically allowed regions,
we show only the 10 TeV case.
Several key trends emerge: $\dmre$ decreases with increasing $\mzd$, as discussed before. For a given $\mzd$, higher $\sqrt{s}$ yields larger $\dmre$, suggesting that the 3 TeV MuC could be more efficient to probe the heavy $\zd$ within its kinematic reach. In contrast, $\dmee$ increases with $\mzd$, reflecting the degrading resolution for higher-energy electron pairs. Interestingly, $\dmee$ and $\dmre$ intersect around $\mzd \simeq \sqrt{s}/2$, suggesting a natural division in the mass range where one observable may outperform the other.

In summary, the dependence of $\dmre$ and $\dmee$ on both $\mzd$ and $\sqrt{s}$ underscores the need for mass-dependent selection strategies when searching for heavy dark $Z$ at a multi-TeV MuC. By applying appropriately tuned mass windows for different $\mzd$ hypotheses, we can substantially enhance signal sensitivity. These optimized mass window selections will play a pivotal role in the full signal-to-background analysis presented in the following section.

\section{Signal and Background Analysis}
\label{sec-analysis}

In Section \ref{sec-review},
we identified two efficient discovery channels for the heavy $\zd$ at multi-TeV MuCs:
\begin{align}
\label{eq-jjX}
\mmu &\to \zd (\to jj X) \gm,
\\ \label{eq-ee}
\mmu &\to \zd (\to \ee) \gm.
\end{align}
This section presents a comprehensive analysis of these channels, including background processes, simulation methodologies, and sensitivity projections.

For the multi-TeV MuC configuration, we consider the following three configurations~\cite{Han:2020pif}:
\bit
\item $\sqrt{s}=3\tev$ with $\lumtot=1\iab$;
\item $\sqrt{s}=6\tev$ with $\lumtot=4\iab$;
\item $\sqrt{s}=10\tev$ with $\lumtot=10\iab$.
\eit
The increase in luminosity with c.m.~energy is attributed to several factors. At higher $\sqrt{s}$, the larger Lorentz factor leads to more collimated muon beams and longer muon lifetime. Additionally, muon beams become tighter due to much less synchrotron radiation at higher energies~\cite{Chiesa:2020awd}.

\subsection{Background Processes and Simulation Methodology}

We first address a unique challenge of a MuC: beam-induced background (BIB). This background arises from the decay of unstable muons in the beam, producing high-energy electron showers that interact with the detector and beamline material, which is particularly severe in the forward regions of the detector. To mitigate BIB, conical tungsten nozzles have been proposed to absorb most soft BIB particles, thereby restricting forward pseudorapidity coverage. Our analysis addresses the BIB through a two-fold strategy: we consider all relevant scattering processes within $|\eta|<2.5$; we apply a hard cut of $p_T > 100\gev$ on the leading jet or electron to suppress soft BIB particles. This approach allows us to focus on the most significant background contributions without explicitly modeling BIB in our simulations.

The primary hard scattering background processes are $\mu^+\mu^-\to jj \gamma$ for the inclusive dijet mode and $\mu^+\mu^-\to e^+e^- \gamma$ for the $e^+e^-$ mode. The detector coverage within $|\eta|<2.5$ introduces additional background channels for both signal modes, including $\mu^+\mu^-\to W^+W^- \gamma$, $\mu^+\mu^-\to \tau^+\tau^-\gamma$, $\mu^+\mu^- \to t\bar{t}\gamma$, and $\mu^+\mu^-\to ZZ\gamma$.

\begin{table}[h]
  \centering
  \setlength{\tabcolsep}{10pt}
  \renewcommand{\arraystretch}{1.2}
  \begin{tabular}{|c||c|c|c|}
  \toprule
\multicolumn{4}{|c|}{Background cross sections in units of fb}
\\ \midrule
 Processes  & $\sqrt{s}=3\tev$ & $\sqrt{s}=6\tev$ & $\sqrt{s}=10\tev$\\ \hline
 $\mmu\to jj\gm$ & $5.01 \times 10$ & $1.40 \times 10$ &  5.44 \\ \hline
  $\mmu\to \ee\gm$ & $ 9.31$ &  2.75 & 1.10 \\ \hline
  $\mmu\to \ww\gm$ & $ 2.55\times 10 $ & $1.01 \times 10$ & 4.86\\ \hline
    $\mmu\to \ttau\gm$ & 5.37 & 1.63 & $6.67 \times 10^{-1}$ \\ \hline
$\mmu\to \ttop\gm$ & 3.24 & 1.07 & $4.58 \times 10^{-1}$ \\ \hline
$\mmu\to ZZ\gm$ & 2.25 & $8.30 \times 10^{-1}$ & $3.82 \times 10^{-1}$  \\ \bottomrule
  \end{tabular}
  \caption{Background cross sections at the 3 TeV, 6 TeV, and 10 TeV MuCs. We imposed cuts on the photon with $p_T^\gamma > 20\gev$ and $|\eta_\gamma|<2.5$. For the decay products of the heavy $\zd$, we additionally imposed $p_T^{j_1, e_1} > 100\gev$, $p_T^{j_2, e_2} > 20\gev$, and $|\eta_{j,e}| < 2.5$.}
  \label{tab-background-xsec}
\end{table}

To assess these background processes, we performed parton-level calculations using {\small\sc MadGraph5-aMC@NLO}~\cite{Alwall:2011uj} version 3.5.1. \autoref{tab-background-xsec} presents the resulting cross sections at 3 TeV, 6 TeV, and 10 TeV MuCs. For each background process at a given $\sqrt{s}$, we generated $5\times 10^5$ events. We apply selection cuts requiring $p_T^\gamma > 20\gev$ and $|\eta_\gamma| < 2.5$ for photons, along with $p_T^{j_1, e_1} > 100\gev$, $p_T^{j_2, e_2} > 20\gev$, and $|\eta_{j,e}| < 2.5$ for the leading and subleading jets or electrons (ordered by descending $p_T$).

The results indicate that cross sections decrease with increasing c.m.~energy across all background processes. The $\mu^+\mu^- \to jj \gamma$ process exhibits the highest cross section at all collision energies. Although $\mu^+\mu^- \to W^+W^- \gamma$ has the second-largest cross section, its contribution is to be suppressed by the branching ratios of $W\to jj$ or $W\to e\nu$. For the $e^+e^-$ mode, $\mu^+\mu^- \to e^+e^-\gamma$ is the dominant background process. The $\mu^+\mu^- \to \tau^+\tau^-\gamma$ process has a slightly smaller cross section than $\mu^+\mu^- \to e^+e^-\gamma$ due to the heavier tau mass. The $\mu^+\mu^- \to t\bar{t}\gamma$ and $\mu^+\mu^- \to ZZ\gamma$ processes have relatively low cross sections.

Building on these parton-level insights, we implemented a comprehensive multi-step simulation process for both signal and background events at the detector level. The process begins with parton-level event generation using {\small\sc MadGraph5-aMC@NLO}, followed by parton showering and hadronization with {\small\sc PYTHIA} version 8.307~\cite{Bierlich:2022pfr}. For detector simulation, we employed {\small\sc DELPHES}. Jet reconstruction was performed using the inclusive Valencia algorithm~\cite{Boronat:2014hva,Boronat:2016tgd} with a jet radius of $R=0.5$ and $p_T>20\gev$ within the {\small\sc FastJet} framework~\cite{Cacciari:2011ma}. The Valencia algorithm is particularly well-suited for high-energy lepton collider environments due to its effectiveness in handling ISR and BIBs through its beam jet concept.

\subsection{Inclusive Two-Jet Signal Analysis}

We first analyze the inclusive dijet mode, $\mmu \to \zd (\to jj X) \gm$, implementing the following basic selection criteria for both signal and background events:
\begin{itemize}
\item $N_j \geq 2$ and $N_\gm \geq 1$ with $p_T^{j,\gm}>20 \gev$ and $|\eta_{j,\gm}| < 2.5$;
\item $p_T^{j_1} > 100\gev$;
\item Veto leptons that have $p_T > 20\gev$  and $|\eta| < 2.5$, i.e., $N_e = N_\mu = 0$.
\end{itemize}
The lepton veto ensures orthogonality between the dijet and $e^+e^-$ samples, enabling independent analysis of the two channels.

\begin{figure}[!t]
  \centering
  \includegraphics[width=0.48\textwidth]{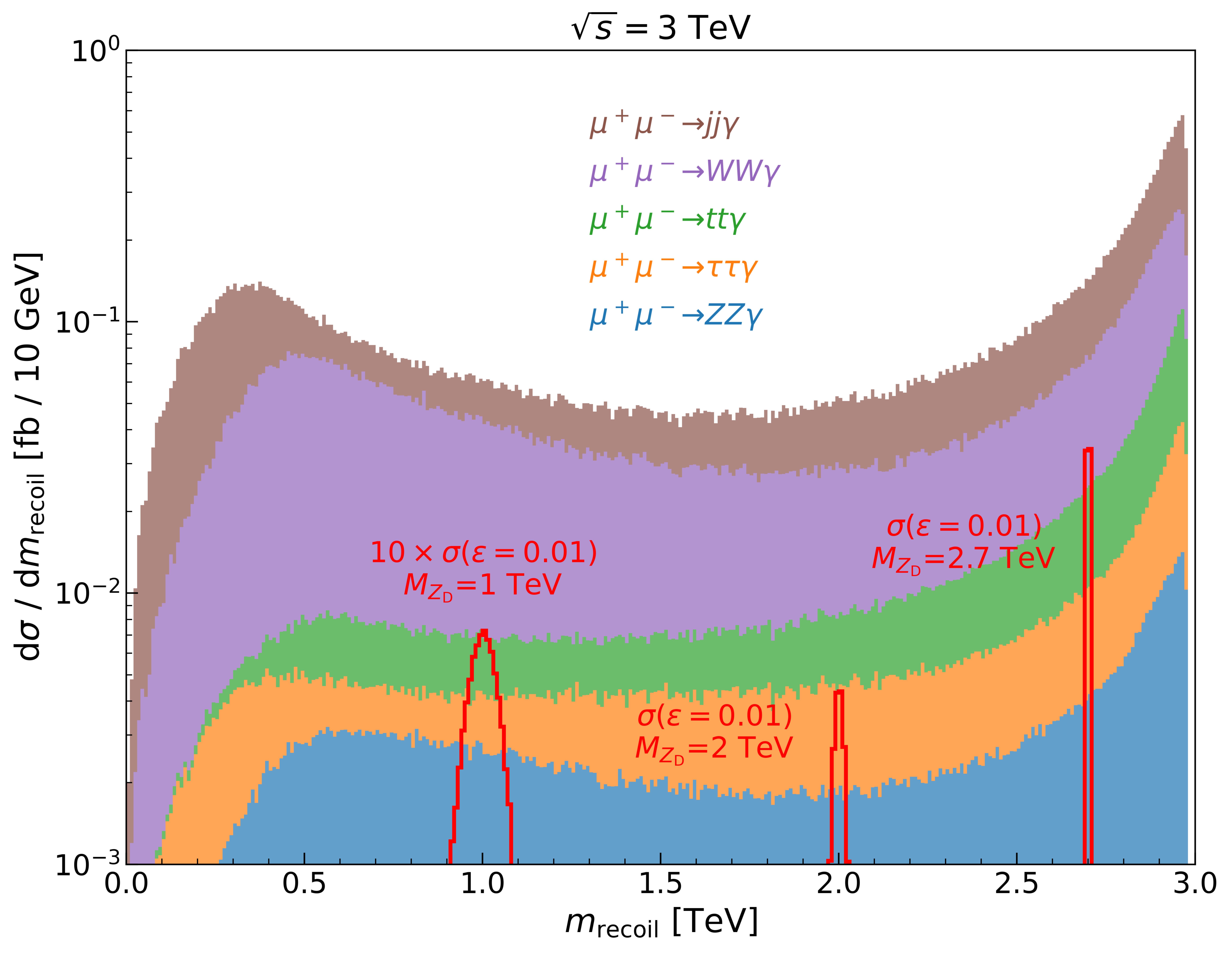}
  \includegraphics[width=0.48\textwidth]{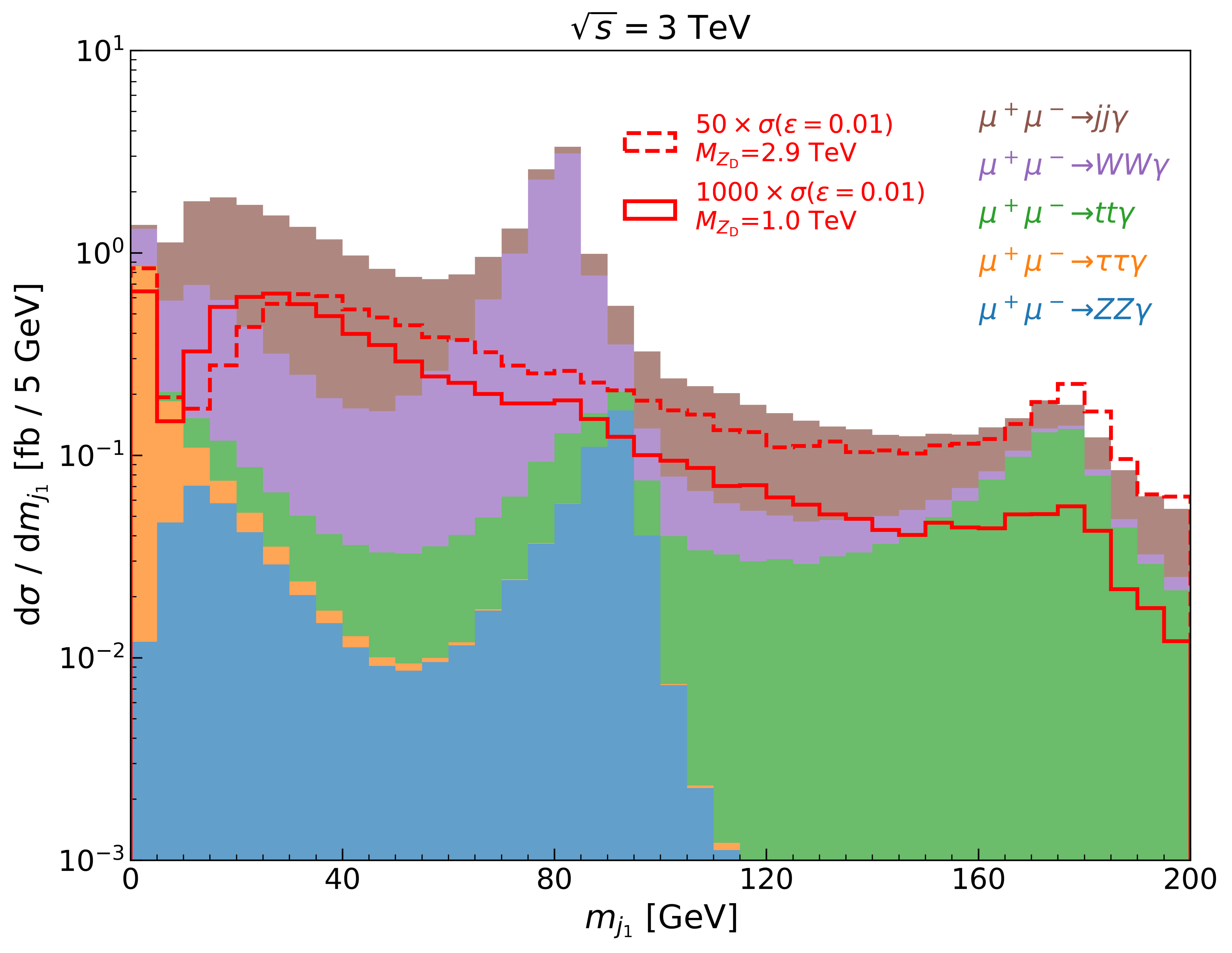}
  \vspace{-0.3cm}
  \caption{Distributions of the recoil mass of the photon, $\mre$, (left) and the invariant mass of the leading jet, $m_{j_1}$, (right), for the signal and backgrounds at the 3 TeV MuC after basic selection. For the signal (red solid lines) with $\ves=0.01$, we consider $\mzd=1\tev$, $2\tev$, and $2.7\tev$ in the $\mre$ distribution, and $\mzd=1\tev$ (solid) and $2.9\tev$ (dashed) in the $m_{j_1}$ distribution. Backgrounds are shown as stacked histograms for direct comparison.
}
  \label{fig-kin-jjX}
\end{figure}

To optimize background suppression in the inclusive dijet mode, we examine two key kinematic observables: the photon recoil mass $\mre$ and the leading jet invariant mass $m_{j_1}$. \autoref{fig-kin-jjX} displays these distributions for both signal and backgrounds at the 3 TeV MuC after applying the basic selection. For the signal benchmarks, we set $\varepsilon=0.01$ and examine $\mzd$ values of 1, 2, and 2.7 TeV in the $\mre$ distribution, while considering 1 and 2.9 TeV for the $m_{j_1}$ distribution. The background contributions are presented as stacked histograms to facilitate direct comparison with the signal.

The $\mre$ distribution reveals distinct features between signal and background events. The signal shows a characteristic narrow peak that becomes sharper for higher $\mzd$ values. In contrast, the background distributions remain nearly flat within the signal peak region. This clear distinction suggests an effective strategy: applying an optimized $\mre$ cut around the signal peak.

The $m_{j_1}$ distribution provides complementary discrimination power. For signal events, we observe substantial nonzero $m_{j_1}$ values that increase with $\mzd$, as shown by comparing the distributions for $\mzd=1\tev$ and $2.9\tev$. This correlation arises from the characteristic behavior of high-energy jets, where the differential cross-section follows $d\sigma/d m_j^2 \sim \alpha_s/m_j^2 \ln (E_j^2/m_j^2)$~\cite{Dasgupta:2012hg}. While following this trend, the signal's $m_j$ distribution remains relatively smooth. In contrast, the background $m_{j_1}$ distributions exhibit distinctive peaks near $m_W$, $m_Z$, and $m_t$. These peaks originate from highly collimated decay products ($R < 0.4$) reconstructed as single jets from $W$, $Z$, and top quark decays, respectively. Among the background processes, the $\mu^+\mu^- \to W^+W^-\gamma$ channel produces the most prominent peak at $m_{j_1} \sim m_W$. Therefore, vetoing events with $m_{j_1}$ in the vicinity of $m_W$ effectively suppresses this background contribution.

\begin{table}[h]
  \centering
  \footnotesize
  \setlength{\tabcolsep}{3pt}
  \renewcommand{\arraystretch}{1.3}
  \begin{tabular}{|c||c||c|c|c|c||c|}
  \toprule
  \multicolumn{7}{|c|}{Cut-flow of cross sections (fb) for $\mu^+ \mu^- \rightarrow \zd (\to jjX) \gm $ at a 3 TeV MuC with $\mathcal{L}_\text{tot}=1\iab$ }\\ \midrule
  \multicolumn{7}{|c|}{$\mzd=1\tev \;\; \& \;\;  \ves =0.01$ } \\ \midrule
Cut & $\zd(\to jj)\gm$ & $jj\gm$ & ~$W^+ W^- \gm$~ &  ~$\ttau\gm$~ & ~$\ttop\gm$~  & $\mathcal{S}_{1\,\text{ab}^{-1}}$\\ \hline
Basic & $7.69 \times 10^{-3}$ & $1.42\times 10$ & $1.15 \times 10$ & 1.08 & 1.86  & $4.47\times 10^{-2}$ \\ \hline
$|m_{j_1}-m_W|>20\gev$ & $6.34 \times 10^{-3}$ & $1.20 \times 10$ & 3.75 & 1.08  & 1.51 & $4.64\times 10^{-2}$ \\ \hline
$|\mre-\mzd|<2\dmre$ & $3.47 \times 10^{-3}$ & $ 9.37 \times 10^{-2}$ & $ 8.47\times 10^{-2}$ & $1.01 \times 10^{-2}$ & $1.29 \times 10^{-2}$ & $2.40 \times 10^{-1}$ \\ \midrule
 \multicolumn{7}{|c|}{$\mzd=2.7\tev \;\; \& \;\;  \ves =0.01$ } \\ \midrule
$|\mre-\mzd|<2\dmre$ & $3.12 \times 10^{-2}$ & $ 3.32 \times 10^{-2}$ & $ 6.22\times 10^{-3}$ & $4.20 \times 10^{-3}$ & $7.72 \times 10^{-3}$ & 3.96 \\ \bottomrule
  \end{tabular}
  \caption{Cut-flow of cross sections in units of fb for the signal $\mu^+ \mu^- \rightarrow \zd(\to jj X)\gm $ with $\ves=0.01$ and $\mzd=1\tev$ and $2.7\tev$ at the 3 TeV MuC. The significance $\mathcal{S}_{1\, \text{ab}^{-1}}$ is calculated assuming an integrated luminosity of 1 ab$^{-1}$. Note that $\dmre$ and $\dmee$ values depend on $\mzd$. }
  \label{tab-cutflow-jjX-3TeV}
\end{table}

To quantify the effectiveness of our selection strategy, we present the cut-flow of cross sections for $\mu^+ \mu^- \rightarrow \zd (\to jjX) \gamma$ at a 3 TeV MuC in \autoref{tab-cutflow-jjX-3TeV}. This table includes the signal significance $\mathcal{S}$, calculated for a total integrated luminosity of $\mathcal{L}_\text{tot}=1\iab$, using the following formula~\cite{Cowan:2010js}:
\begin{equation}
\mathcal{S} = \sqrt{2\left[\left(n_s+n_b \right)\ln\left(\frac{n_s +n_b }{n_b}\right)-n_s\right]},
\end{equation}
where $n_s$ and $n_b$ denote the number of signal and background events, respectively.

For the detailed cut-flow analysis, we examine two $\mzd$ hypotheses with $\ves=0.01$: $\mzd=1\tev$ and $2.7\tev$. Since the cut-flow results up to the $m_{j_1}$ cut are nearly identical for both masses, we present the complete cut-flow only for $\mzd=1\tev$, showing just the final result for $\mzd=2.7\tev$. We include all background processes except $\mmu\to ZZ\gm$, which we exclude due to its negligible cross section.

The cut-flow results in \autoref{tab-cutflow-jjX-3TeV} demonstrate the progressive improvement of our signal selection. After basic selection, the $jj\gamma$ and $W^+W^-\gamma$ backgrounds dominate, yielding a low initial signal significance of approximately 0.02. The $W$ veto, achieved by requiring $|m_{j_1}-m_W|>20\gev$, proves moderately effective, suppressing about 67\% of the $W^+W^-\gamma$ background. The most substantial improvement comes from the $\mre$ cut, which significantly suppresses all backgrounds while largely preserving the signal, demonstrating the power of the recoil mass as a discriminating variable.

Nevertheless, our final selection achieves only about $0.24\sigma$ significance for $\mzd=1\tev$ and $\ves=0.01$, far below the detection level. In contrast, the heavier $\zd$ case ($\mzd=2.7\tev$) reaches a dramatic enhancement in significance to $4.3\sigma$, approaching the discovery threshold. This substantial improvement over the $\mzd=1\tev$ case primarily stems from the difference in $\dmre$: at the 3 TeV MuC, $\dmre =16.3$ GeV for $\mzd=1\tev$, compared to $\dmre=1.66$ GeV for $\mzd=2.7\tev$. At a given c.m.~energy, therefore, heavier $\zd$ achieves higher discovery significance.

\begin{table}[h]
  \centering
  \footnotesize
  \setlength{\tabcolsep}{3pt}
  \renewcommand{\arraystretch}{1.3}
  \begin{tabular}{|c||c||c|c|c|c||c|}
  \toprule
  \multicolumn{7}{|c|}{Cut-flow of cross sections [fb] for $\mu^+ \mu^- \rightarrow \zd (\to jjX) \gm $ at a 10 TeV MuC with $\mathcal{L}_\text{tot}=10\iab$ }\\ \midrule
  \multicolumn{7}{|c|}{$\mzd=1\tev \;\; \& \;\;  \ves =0.01$ } \\ \midrule
Cut & $\zd(\to jj)\gm$ & $jj\gm$ & ~$W^+ W^- \gm$~ &  ~$\ttau\gm$~ & ~$\ttop\gm$~  & $\mathcal{S}_{10\,\text{ab}^{-1}}$\\ \hline
Basic & $5.09 \times 10^{-4}$ & 1.21 & 1.86 & $1.37 \times 10^{-1}$  & $2.91 \times 10^{-1}$  & $2.69\times 10^{-2}$ \\ \hline
$|m_{j_1}-m_W|>20\gev$ & $4.02 \times 10^{-4}$ & $9.25\times 10^{-1} $ & $5.14\times 10^{-1} $ & $1.33\times 10^{-1} $  & $2.78 \times 10^{-1} $ & $2.93\times 10^{-2}$ \\ \hline
$|\mre-\mzd|<2\dmre$ & $2.28  \times 10^{-4}$ & $ 4.55\times 10^{-2}$ & $ 9.00\times 10^{-2}$ & $4.49 \times 10^{-3}$ & $9.23 \times 10^{-3}$ & $5.85 \times 10^{-2}$ \\ \midrule
 \multicolumn{7}{|c|}{$\mzd=9.7\tev \;\; \& \;\;  \ves =0.01$ } \\ \midrule
$|\mre-\mzd|<2\dmre$ & $1.18 \times 10^{-2}$ & $ 3.18 \times 10^{-3}$ & $ 9.43\times 10^{-4}$ & $3.95 \times 10^{-4}$ & $1.20 \times 10^{-3}$ & 12.4 \\ \bottomrule
  \end{tabular}
  \caption{Cut-flow of the cross sections in units of fb 
  for the signal $\mu^+ \mu^- \rightarrow \zd(\to jj X)\gm $ with $\ves=0.01$ and $\mzd=1\tev,9.7\tev$ at the 10 TeV MuC.   The significance $\mathcal{S}_{10\, {\rm ab}^{-1}}$ is calculated considering an integrated luminosity of 10 ab$^{-1}$. Note that $\dmre$ and $\dmee$ values depend on $\mzd$.}
  \label{tab-cutflow-jjX-10TeV}
\end{table}

To explore the effects of higher c.m.~energy, we extend our analysis to a 10 TeV MuC with a total integrated luminosity of 10 ab$^{-1}$, as presented in \autoref{tab-cutflow-jjX-10TeV}. As in our 3 TeV analysis, we omit the negligible $ZZ\gamma$ background and focus on the remaining significant background processes.

We examine two $\mzd$ hypotheses with $\ves=0.01$: $\mzd=1$ TeV and 9.7 TeV. The results reveal a striking contrast between lighter and heavier $\zd$ scenarios. For $\mzd=1\tev$, both signal and background cross sections decrease compared to the 3 TeV case, resulting in a reduced final significance of only $0.058\sigma$. This decline illustrates the challenge of detecting lighter $\zd$ at higher collision energies. 
In contrast, the $\mzd=9.7\tev$ case demonstrates the unique potential of high-energy MuCs for heavy $\zd$ searches, achieving a dramatic enhancement in significance to $12.4\sg$. This remarkable improvement arises from three factors: the enhanced signal cross section after basic selection (approximately $2.44 \times 10^{-2}$ fb), the much narrower $\dmre$ of about 1.77 GeV enabling highly efficient background suppression, and the ten times higher integrated luminosity ($10\iab$) compared to the 3 TeV MuC. These results underscore the complementarity of different collision energies in exploring the $\mzd$ parameter space, with higher-energy MuCs excelling at probing heavier $\zd$ masses.

\subsection{Analysis of the $\zd \to e^+e^-$ Channel}

We now analyze the $\ee$ mode, $\mmu \to \zd (\to \ee) \gm$, beginning with the following basic selection criteria:
\bit
\item $N_e = 2$ and $N_\gm \geq 1$ with $\pt>20\gev$ and $|\eta_{e,\gm}|<2.5$;
\item $\pt^{e_1} >100\gev$;
\item Veto jets or muons that have $p_T > 20\gev$ and $|\eta| < 2.5$, i.e., $N_j=N_\mu=0$.
\eit
The jet and muon veto ensures that our $e^+e^-$ sample remains distinct from the inclusive dijet mode.

\begin{figure}[!t]
  \centering
  \includegraphics[width=0.49\textwidth]{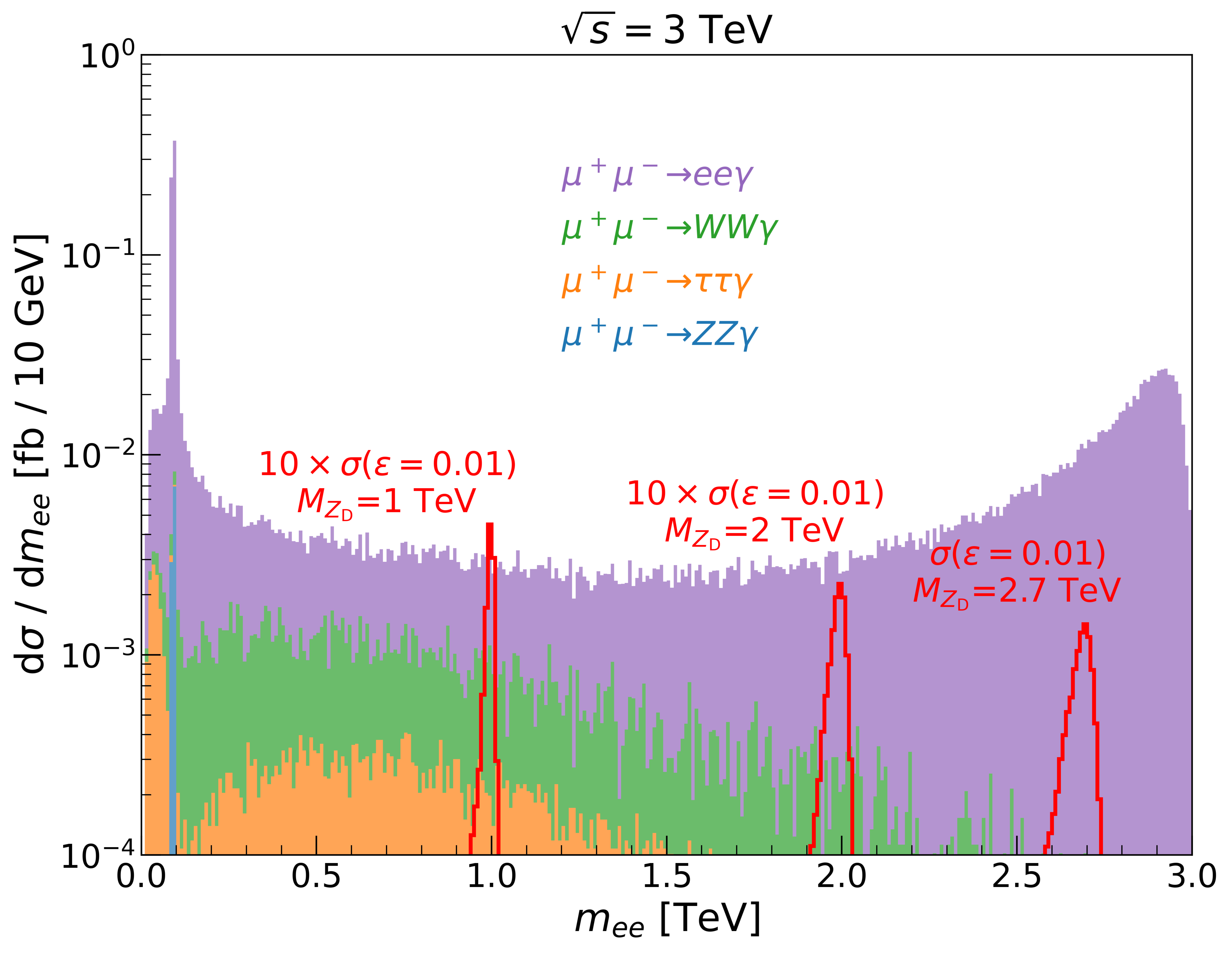}
  \includegraphics[width=0.49\textwidth]{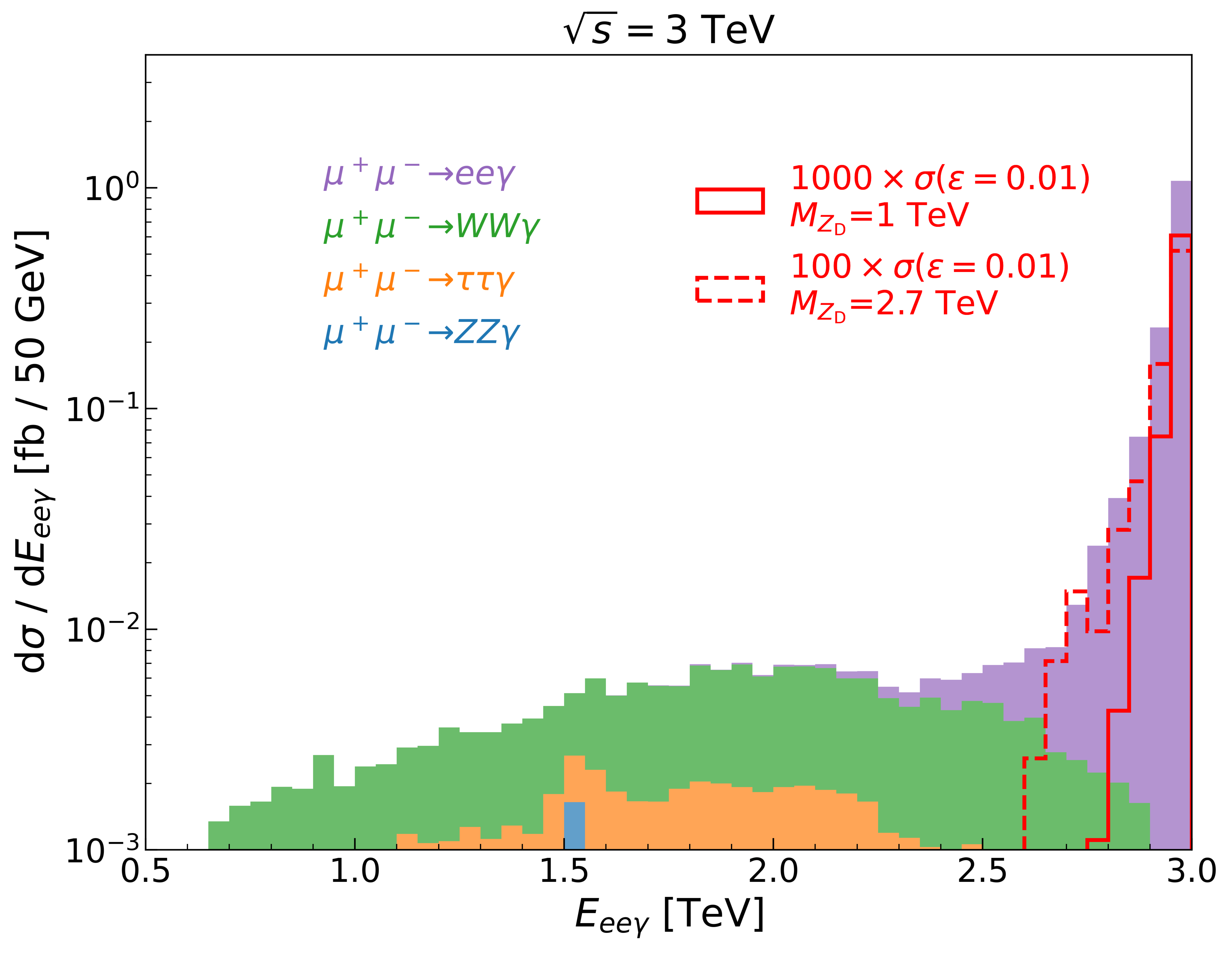}
  \vspace{-0.3cm}
  \caption{Distributions of the invariant mass of the $e^+e^-$ pair (left) and the total energy of the $e^+e^-\gamma$ system, $E_{ee\gamma}$, (right) for the signal and backgrounds at the 3 TeV MuC after the basic selection. For the signal (red solid lines), we set $\varepsilon=0.01$ and consider $\mzd=1,2,2.7\tev$ for the $m_{ee}$ distribution, and $\mzd=1,2.7\tev$ for the $E_{ee\gamma}$ distribution. Backgrounds are shown as stacked histograms for direct comparison.
}
  \label{fig-kin-ee}
\end{figure}

To optimize our background suppression strategy in the $e^+e^-$ mode, we examine two key kinematic observables: the invariant mass of the electron-positron pair ($m_{ee}$) and the total energy of the $e^+e^-\gamma$ system ($E_{ee\gamma}$). \autoref{fig-kin-ee} presents these distributions for both signal and backgrounds at the 3 TeV MuC after applying the basic selection. Note that the recoil mass distribution is nearly identical to that in the $jjX$ mode, as shown in \autoref{fig-kin-jjX}.
In \autoref{fig-kin-ee}, we consider the signal with $\varepsilon=0.01$ for three mass hypotheses
($\mzd=1,2,2.7\tev$) in the $m_{ee}$ distribution and two mass hypotheses ($\mzd=1, 2.7 \tev$) in the $E_{ee\gamma}$ distribution.

The $m_{ee}$ distribution exhibits clear signal features: a resonance peak whose width increases with $\mzd$. In contrast, the background $m_{ee}$ distribution remains smooth across each signal peak region, suggesting that a narrow $m_{ee}$ cut would effectively suppress the background while maintaining signal efficiency.

The $E_{ee\gamma}$ distribution reveals another distinctive feature: both the signal and the irreducible background ($\mu^+ \mu^- \to e^+e^-\gamma$) peak at $\sqrt{s}$. This behavior occurs because these processes produce only $e^+e^-\gamma$ in the final state, retaining most of the collision energy despite small losses to FSR. In contrast, reducible background processes, which produce additional particles, exhibit significant energy loss in the $e^+e^-\gamma$ system. This characteristic suggests that requiring $E_{ee\gamma} > 0.9\sqrt{s}$ would efficiently suppress all the reducible backgrounds.

\begin{table}[h]
  \centering
  \footnotesize
  \setlength{\tabcolsep}{3pt}
  \renewcommand{\arraystretch}{1.3}
  \begin{tabular}{|c||c|c||c|c||c|c|}
  \toprule
  \multicolumn{7}{|c|}{Preliminary cut-flow of cross sections [fb] for $\mu^+ \mu^- \rightarrow \zd (\to \ee ) \gm $ at a 3 TeV MuC with $\mathcal{L}_\text{tot}=1\iab$ }\\ \midrule
  \multirow{2}{*}{Cut}  &  \multicolumn{2}{c||}{Backgrounds} &  \multicolumn{2}{c||}{$\mzd=1\tev$}  & \multicolumn{2}{c|}{$\mzd=2.7\tev$} \\ \cline{2-7}
  & ~$\ee\gm$~ & ~$\ww\gm$~ & ~$\zd(\to\ee) \gm$~ &  $\mcs_{1\,\text{ab}^{-1}}$ & ~$\zd(\to\ee) \gm$~ &  $\mcs_{1\,\text{ab}^{-1}}$\\ \hline
Basic & 2.14 & ~$1.32\times 10^{-1}$~ & $1.10 \times 10^{-3}$ & ~$2.28 \times 10^{-2}$~  & $1.07 \times 10^{-2}$  & $2.22\times 10^{-1}$ \\ \hline
~$E_{ee\gm}>0.9\sqrt{s}$~ & 2.11 & $7.85\times 10^{-3} $ & $1.09\times 10^{-3} $ & $2.37\times 10^{-2} $  & $1.06\times 10^{-2} $ & $2.30\times 10^{-1}$  \\ \bottomrule
  \end{tabular}
  \caption{Preliminary cut-flow of cross sections (fb) for the signal $\mu^+ \mu^- \rightarrow Z_D (\to e^+e^-) \gamma$ with $\varepsilon=0.01$ and $\mzd=1\ \text{TeV}, 2.7\ \text{TeV}$ at the 3 TeV MuC. The significance $\mathcal{S}_{1,\text{ab}^{-1}}$ is calculated assuming an integrated luminosity of 1 ab$^{-1}$.}
  \label{tab-cutflow0-ee-3TeV}
\end{table}

To quantify the impact of the $E_{ee\gamma}$ cut, we present a preliminary cut-flow analysis in \autoref{tab-cutflow0-ee-3TeV} for $\mu^+ \mu^- \rightarrow Z_D (\to e^+e^-) \gamma$ at a 3 TeV MuC with $\lumtot=1\ \text{ab}^{-1}$. We examine two benchmark points with $\varepsilon=0.01$: $\mzd=1\ \text{TeV}$ and $\mzd=2.7\ \text{TeV}$. After basic selection, the irreducible background $\mu^+ \mu^- \to e^+e^-\gamma$ dominates, exceeding the signal by factors of approximately 1000 and 100 for $\mzd=1\ \text{TeV}$ and $\mzd=2.7\ \text{TeV}$, respectively. The $E_{ee\gamma} > 0.9\sqrt{s}$ cut proves highly effective in suppressing the $\mu^+ \mu^- \to W^+W^-\gamma$ background, reducing its cross section to approximately 6\% of its value before applying this cut.

\begin{table}[h]
  \centering
  \footnotesize
  \setlength{\tabcolsep}{3pt}
  \renewcommand{\arraystretch}{1.3}
  \begin{tabular}{|c||c|c||c|c||c|c|}
  \toprule
  \multicolumn{7}{|c|}{Final selection results for $\mu^+ \mu^- \rightarrow \zd (\to \ee ) \gm $ at a 3 TeV MuC with $\mathcal{L}_\text{tot}=1\iab$ }\\ \midrule
  & \multicolumn{2}{c||}{$\mre$ cut } & \multicolumn{2}{c||}{$\mee$ cut }  & \multicolumn{2}{c|}{combined} \\ \hline
   & $\sg_{\rm sig}$ [fb] & $\mcs_{1\,\text{ab}^{-1}}$ &    $\sg_{\rm sig}$ [fb] & $\mcs_{1\,\text{ab}^{-1}}$   &  $\sg_{\rm sig}$ [fb] & $\mcs_{1\,\text{ab}^{-1}}$ \\ \hline
$\mzd=1\tev$ & $6.01\times 10^{-4}$ & $1.69 \times 10^{-1}$ & $ 6.88 \times 10^{-4}$ & $3.75\times 10^{-1}$
&$3.79 \times 10^{-4}$ & $3.04 \times 10^{-1}$ \\ \hline
$\mzd=2.7\tev$ & $5.91\times 10^{-3}$ & 2.12& $ 5.71 \times 10^{-3}$ & $7.37\times 10^{-1}$ 
 &$3.23 \times 10^{-3}$ & 1.61 \\ \bottomrule
  \end{tabular}
  \caption{Signal cross sections and significances for $\mu^+ \mu^- \rightarrow Z_D (\to e^+e^-) \gamma$ at a 3 TeV MuC with $\mathcal{L}_\text{tot}=1$ ab$^{-1}$, shown for $\mzd=1\ \text{TeV}$ and $\mzd=2.7\ \text{TeV}$. We fixed $\varepsilon=0.01$. Results are presented after basic selection, $E_{ee\gamma} > 0.9\sqrt{s}$ cut, and three final selection criteria: (1) $\mre$ cut, (2) $m_{ee}$ cut, and (3) combined cut applying both conditions.
}
  \label{tab-final-cutflow-ee-3TeV}
\end{table}

With the preliminary cuts established, a crucial question arises: should we apply a cut on $\mre$, $m_{ee}$, or both? The choice is non-trivial since the optimized mass window cuts for $\mre$ and $m_{ee}$ show opposite behaviors with respect to $\mzd$. For a comparative analysis of these final selection strategies, \autoref{tab-final-cutflow-ee-3TeV} presents the signal significance of three different final selections: (1) $\mre$ cut only, (2) $m_{ee}$ cut only, and (3) combined cut applying both conditions. The basic selection and $E_{ee\gamma} > 0.9\sqrt{s}$ cut from \autoref{tab-cutflow0-ee-3TeV} have been applied.

For lighter $Z_D$ ($\mzd=1\ \text{TeV}$), the $m_{ee}$ cut alone proves most efficient, yielding a significance 2.2 times that achieved with the $\mre$ cut alone. This aligns with the opposite behavior of $\Delta \mre$ and $\Delta m_{ee}$ with respect to $\mzd$, as illustrated in \autoref{fig-Delta-Mass}. Notably, the $m_{ee}$ cut also outperforms the combined cut, achieving a significance about 20\% higher.

For heavier $Z_D$ ($\mzd=2.7\ \text{TeV}$), the situation reverses. The $\mre$ cut alone is most effective, achieving a significance of approximately 2.41 for $\varepsilon=0.01$. This value is 2.9 times that obtained with the $m_{ee}$ cut alone and 1.4 times that achieved with the combined cut.

\subsection{Combined Results}

In this final subsection, we present comprehensive results of $2\sigma$ and $5\sigma$ sensitivity projections in the $(\mzd,\ves)$ parameter space by combining the inclusive dijet and $\ee$ modes. Our analysis is based on detailed detector-level simulations, as described in the previous sections. To systematically explore the parameter space, we selected benchmark points with $\ves=0.01$: 12 values of $\mzd$ for the 3 TeV MuC, another 12 for the 6 TeV MuC, and 20 values for the 10 TeV MuC. For each $\mzd$ benchmark point, we generated $5 \times 10^4$ signal events. Since $\ves$ affects only the total signal rate while preserving the distribution shape, we can efficiently explore the full $(\mzd,\ves)$ parameter space by scaling the signal cross section according to $\ves^2$.

 \begin{figure}[!t]
  \centering
  \includegraphics[width=0.48\textwidth]{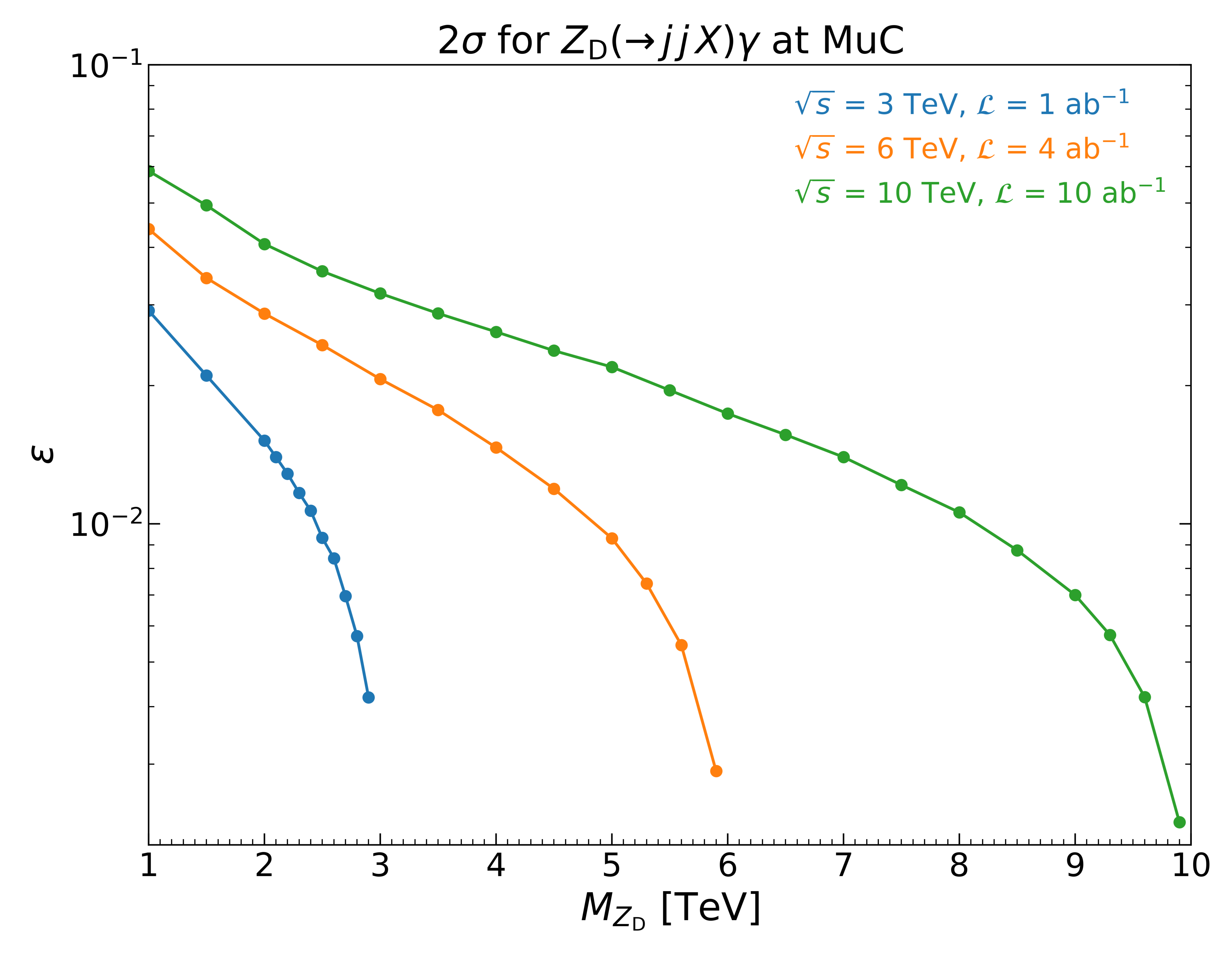}
  \includegraphics[width=0.48\textwidth]{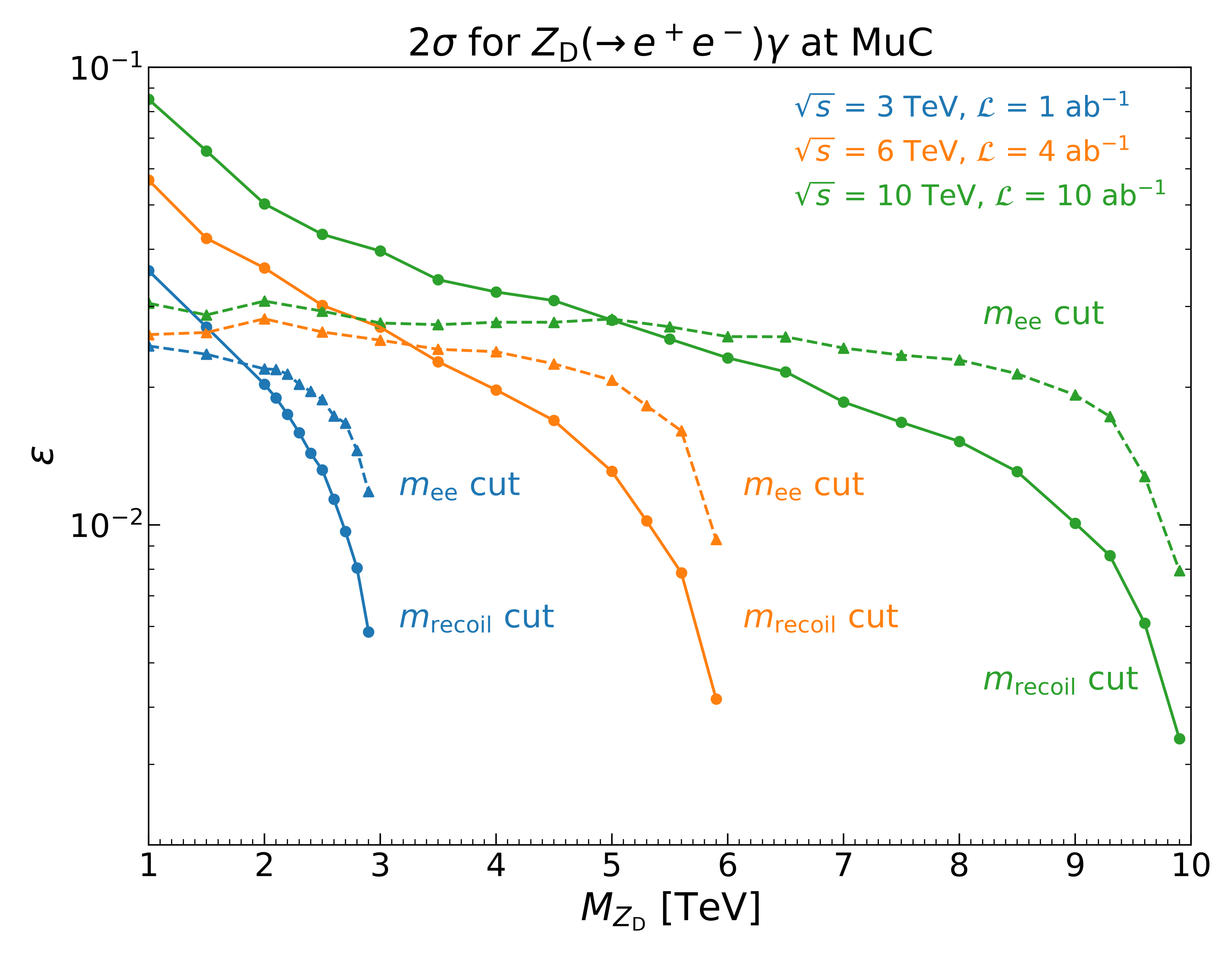}
  \vspace{-0.3cm}
  \caption{$2\sigma$ sensitivity contours for detecting a dark $Z$ boson at different MuC energies through the processes $\mu^+ \mu^- \to Z_D (\to jjX) \gamma$ (left) and $\mu^+ \mu^- \to Z_D (\to e^+e^-) \gamma$ (right). The results for the 3 TeV MuC with $\mathcal{L}_\text{tot}=1\ \text{ab}^{-1}$, 6 TeV MuC with $\mathcal{L}_\text{tot}=4\ \text{ab}^{-1}$, and 10 TeV MuC with $\mathcal{L}_\text{tot}=10\ \text{ab}^{-1}$ are shown in blue, orange, and green, respectively. For the $e^+e^-$ mode, solid lines denote results based on the $\mre$ cut, while dashed lines denote those based on the $\mee$ cut.
}
  \label{fig-2sigma-jjX-ee}
\end{figure}

We first present the $2\sigma$ sensitivity contours in \autoref{fig-2sigma-jjX-ee}, shown separately for the $\zd\to jjX$ mode (left panel) and $\zd\to \ee$ mode (right panel). We examine three MuC configurations: $\sqrt{s}=3\tev$ with $\lumtot=1\iab$ (blue), $\sqrt{s}=6\tev$ with $\lumtot=4\iab$ (orange), and $\sqrt{s}=10\tev$ with $\lumtot=10\iab$ (green).

For the $jjX$ mode, the results are based on the cut-flow outlined in \autoref{tab-cutflow-jjX-3TeV}. Our analysis reveals two universal trends. First, the $\ves$ sensitivity limit as a function of $\mzd$ exhibits a consistent shape across all c.m.~energies, improving rapidly with increasing $\mzd$. This enhancement stems from the increasing cross section and tighter $\mre$ cuts enabled by heavier $\mzd$. Second, the $\ves$ sensitivity extends to higher $\mzd$ values as the collision energy increases, a direct consequence of the expanded kinematic phase space.

For the $e^+e^-$ mode, we present two sets of results: one based on the $\mre$ cut (solid lines) and another on the $\mee$ cut (dashed lines). Both cuts yield similar trends in $\ves$ sensitivity, improving as $\mzd$ increases. The $\mee$ cut enhances $\ves$ sensitivity for heavier $\mzd$, even though our optimized cut on the $\mee$ distribution becomes weaker. This is primarily due to the increased signal cross section for heavier $\zd$ masses, as shown in \autoref{fig-xsec}.

While the $\ves$ sensitivity as a function of $\mzd$ shows similar behavior for both $\mre$ and $\mee$ cuts, their actual values differ significantly across the $\mzd$ range.
 For lighter $\mzd$ values, the $m_{ee}$ selection consistently outperforms the $\mre$ selection across all collision energies. For instance, at $\mzd=1\ \text{TeV}$ in the 3 TeV MuC, the $2\sigma$ sensitivity limit on $\ves$ improves from 0.036 with the $\mre$ selection to 0.025 with the $m_{ee}$ selection. 
 This advantage becomes more pronounced at higher energies; for example, at 10 TeV, the $m_{ee}$ selection achieves a sensitivity limit of 0.030, about 2.8 times better than the $\mre$ selection limit of 0.085.
  These results demonstrate the superiority of the $m_{ee}$ selection for lighter $\mzd$.
Conversely, for heavier $\mzd$ values, the $\mre$ selection becomes more effective. 

The $2\sigma$ sensitivity contours for the $m_{ee}$ and $\mre$ selections intersect near $\mzd \simeq \sqrt{s}/2$. Based on this feature, we implement the $m_{ee}$ selection for $\mzd < \sqrt{s}/2$ and the $\mre$ selection for $\mzd > \sqrt{s}/2$ in the $e^+e^-$ mode.

Finally, we combine the two processes, $\mmu\to \zd(\to jjX)\gm$ and $\mmu\to \zd(\to \ee)\gm$. Statistical independence between the two measurements is ensured by the lepton veto in the inclusive dijet signal and the jet and muon veto in the $\ee$ mode. Assuming Gaussian distributions, we approximate the combined significance as $\mcs_\text{tot} \simeq \sqrt{\mcs_{jjX}^2 + \mcs_{ee}^2}$, where $\mcs_{jjX}$ ($\mcs_{ee}$) is the significance from the $jjX$ ($\ee$) mode. This approach enables efficient computation of the $2\sigma$ and $5\sigma$ sensitivity limits across the entire parameter space. While a Poisson distribution would be more appropriate given the low event count, our explicit calculations for several benchmark points show that Fisher's method for combining $p$-values from the Poisson distribution yields results nearly identical to the Gaussian approximation, differing by only a few percent.

 \begin{figure}[!t]
  \centering
  \includegraphics[width=0.7\textwidth]{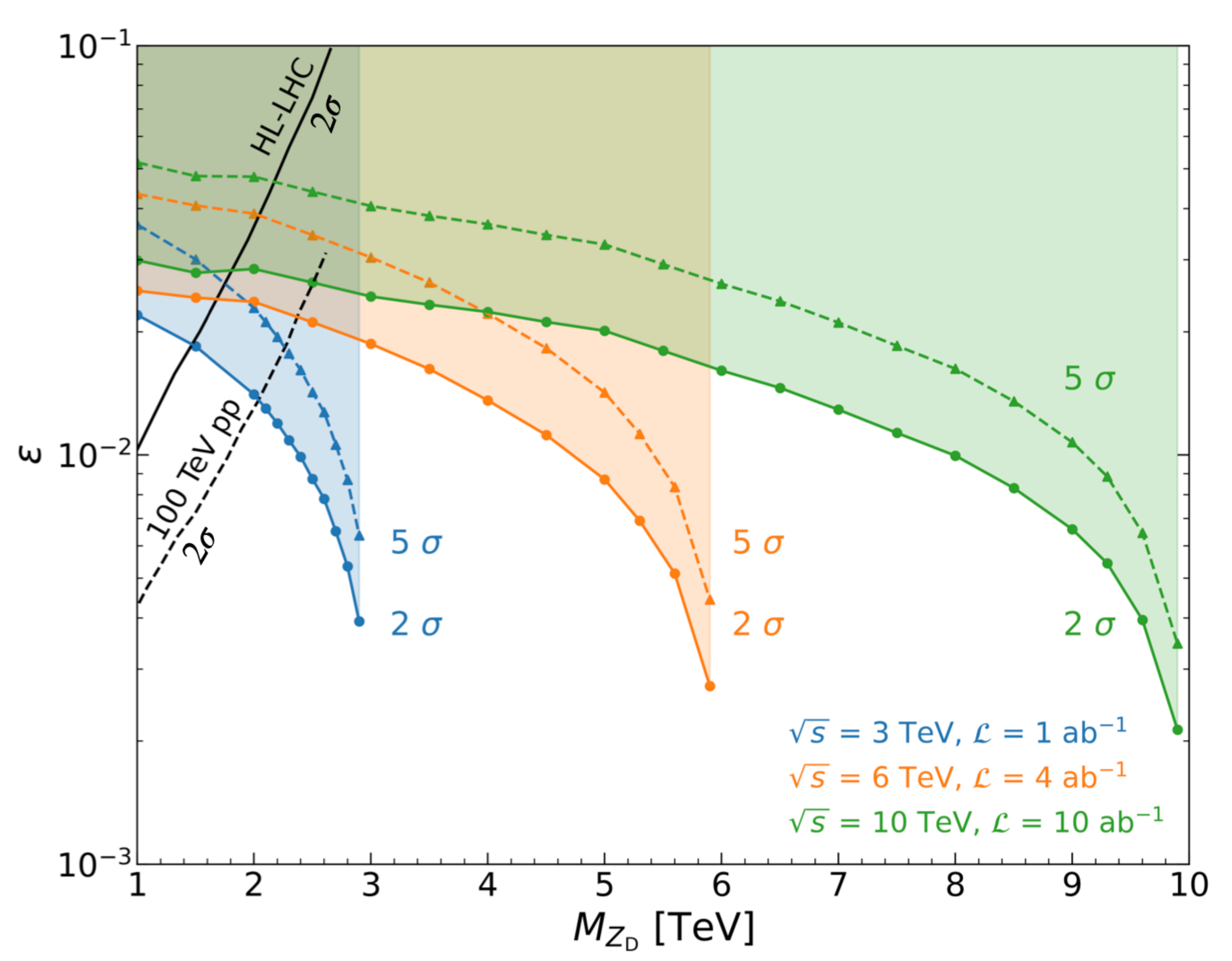}
  \caption{$2\sigma$ (solid lines) and $5\sigma$ (dashed lines) sensitivity contours for detecting a dark $Z$ boson at different MuC energies, combining the processes $\mmu\to \zd(\to jjX)\gm$ and $\mmu\to \zd(\to \ee)\gm$. The results for the 3 TeV MuC with $\lumtot=1\iab$, 6 TeV with $\lumtot=4\iab$, and 10 TeV with $\lumtot=10\iab$ are shown in blue, orange, and green, respectively. The $2\sigma$ sensitivity limit of $\ves$ at the HL-LHC ($\sqrt{s}=14\tev$, $\lumtot=3\iab$) and at a 100 TeV $pp$ collider ($\lumtot=3\iab$)~\cite{Curtin:2014cca} are also shown.
}
  \label{fig-final-epsilon-combined}
\end{figure}

\autoref{fig-final-epsilon-combined} presents our final results, combining the inclusive dijet and $e^+ e^-$ modes. The solid and dashed lines show the $2\sigma$ and $5\sigma$ sensitivity contours as a function of $\mzd$, respectively. Results for different MuC configurations are shown in distinct colors: blue for the 3 TeV MuC with $\lumtot=1\iab$, orange for 6 TeV with $\lumtot=4\iab$, and green for 10 TeV with $\lumtot=10\iab$. For comparison, we include the $2\sigma$ sensitivity limits on $\ves$ from previous studies of the HL-LHC ($\sqrt{s}=14\tev$, $\lumtot=3\iab$) and a 100 TeV $pp$ collider ($\lumtot=3\iab$)~\cite{Curtin:2014cca}.

Our analysis demonstrates the advantages of combining different decay modes to enhance the sensitivity to the coupling strength $\ves$ of the $\zd$ boson. For $\mzd=1\tev$ at the 3 TeV MuC, while the $\ee$ decay mode alone can probe down to $\ves = 2.46\times 10^{-2}$ at $2\sigma$ significance, the combination of $jjX$ and $\ee$ modes improves this limit to $\ves = 2.19\times 10^{-2}$, representing an 11\% enhancement in sensitivity. At the 6 TeV and 10 TeV MuC, combining the two channels for $\mzd=1\tev$ yields more modest improvements of approximately 3.5\% and 2\%, respectively.

For the heaviest accessible cases ($\mzd =\sqrt{s}-100\gev$) at each collision energy, combining both channels allows us to reach $\ves=3.9\times 10^{-3}$ at the 3 TeV MuC, $\ves=2.7\times 10^{-3}$ at the 6 TeV MuC, and $\ves=2.1\times 10^{-3}$ at the 10 TeV MuC. Relative to the $\ves$ sensitivity limit in the $\ee$ mode alone (see \autoref{fig-2sigma-jjX-ee}), the addition of the inclusive dijet mode improves the sensitivity by approximately 6.4\%, 5.8\%, and 4.8\% at the 3 TeV, 6 TeV, and 10 TeV MuC, respectively.

Although the improvements from combining two channels are modest, they provide meaningful enhancements to our search sensitivity. The consistent gains across different energies and masses demonstrate the value of the multi-channel approach and motivate further refinement of these techniques in future analyses.

Finally, we compare our results with the sensitivity projections for the HL-LHC ($\sqrt{s}=14\tev$, $\lumtot=3\iab$) and a 100 TeV $pp$ collider ($\lumtot=3\iab$)~\cite{Curtin:2014cca}. For $\mzd=1\tev$, the HL-LHC can achieve a sensitivity limit of $\ves \sim 10^{-2}$, while the 100 TeV $pp$ collider reaches $\ves \sim 4.3 \times 10^{-3}$. For this relatively light $\mzd$, both hadron colliders outperform the MuC.

However, as $\mzd$ increases, the MuC demonstrates significant advantages. In comparison with the 3 TeV MuC, the HL-LHC achieves comparable $\ves$ sensitivity at $\mzd=1.5\tev$, while the 100 TeV $pp$ collider reaches similar sensitivity at $\mzd=2\tev$. Beyond these mass points, the sensitivity of both hadron colliders deteriorates rapidly, with their reach limited to $\mzd \simeq 2.7\tev$ for $\ves=0.1$. In contrast, the multi-TeV MuC becomes increasingly effective at probing heavier $\mzd$ values within its kinematic reach. This superior performance persists across different MuC energies, establishing the MuC as a promising facility for exploring heavy dark $Z$ bosons.

\section{Conclusions}
\label{sec-conclusion}

In this study, we have explored the discovery potential of multi-TeV muon colliders (MuCs) for a heavy dark $Z$ boson ($Z_D$)  through the associated production channel $\mu^+\mu^- \to Z_D \gamma$. This production mechanism is central to our approach, as it enables precise $\mzd$ reconstruction through the recoil mass technique with well-measured photon energy.
 We focused on two decay modes, $Z_D \to jjX$ and $Z_D \to e^+e^-$, and demonstrated that MuCs operating at 3, 6, and 10 TeV with integrated luminosities of 1, 4, and 10 ab$^{-1}$, respectively, can significantly improve sensitivity to the kinetic mixing parameter $\varepsilon$ over a broad $Z_D$ mass range—particularly above 1 TeV—compared to hadron colliders.

A key advancement in our analysis lies in the optimized implementation of mass-dependent cuts on both the recoil mass ($\mre$) and the invariant mass of the electron-positron pair ($\mee$). Rather than employing fixed mass windows, we introduced $\mzd$-dependent resolutions $\dmre$ and $\dmee$ to account for the energy-dependent detector response. This consideration is crucial because detector energy resolution deteriorates for more energetic photons and electrons. Since a lighter $\zd$ results in a more energetic photon but less energetic electrons, this approach offers clear advantages: for lighter $\zd$ masses, an $\mee$-based selection achieves superior sensitivity, while for heavier $\zd$, the $\mre$-based selection becomes more effective. Indeed, the stringent $\mre$ cut enables the multi-TeV MuC to attain significantly high $\ves$ sensitivity for heavier $\zd$. For the $\zd \to e^+e^-$ mode, we identified a natural crossover point at $\mzd \simeq \sqrt{s}/2$, where the $\mee$-based selection performs better below this mass and the $\mre$-based selection excels above it. This strategy enhances sensitivity across the entire mass spectrum.

The combination of $jjX$ and $e^+e^-$ decay channels further improves the overall sensitivity. While for $\mzd \lesssim 2\tev$, the sensitivity gains at MuCs are comparable to those at the HL-LHC or a 100 TeV $pp$ collider, our strategy achieves $\varepsilon$ sensitivity down to $\mathcal{O}(10^{-3})$ as $\mzd$ approaches $\sqrt{s}$, substantially outperforming any future hadron colliders. These results demonstrate the fundamental advantages of MuCs—the ability to fully exploit the c.m.~energy and provide cleaner final states—enabling them to surpass hadron collider performance for heavier $Z_D$ masses.

Looking ahead, these findings underscore the promise of MuCs in probing heavy dark sectors that remain inaccessible at hadron colliders. The photon recoil mass technique remains effective even when $\zd$ decays into dark sector particles. As detector simulations, reconstruction algorithms, and statistical tools continue to advance, MuCs will likely achieve even greater sensitivity. These developments establish the multi-TeV MuC as a pivotal facility for deepening our understanding of the dark sector.

\section*{Acknowledgments}
We thank Dr.~Adil Jueid and Dr.~Daohan Wang for useful discussions. 
The work of K.C. is supported by National Science and Technology Council (NSTC) 113-2112-M-007-041-MY3.
The work of P.S., and J.S. is supported by
the National Research Foundation of Korea, Grant No.~NRF-2022R1A2C1007583.
JK supported by a KIAS Individual Grant (PG099201) at Korea Institute for Advanced Study.
The work of S.L. is supported by Basic Science Research Program through the National Research Foundation of Korea(NRF) funded by the Ministry of Education(RS-2023-00274098).

\bibliographystyle{JHEPMod}

\bibliography{dark-photon-MuC}

\end{document}